\documentclass[nolinenumbers, longauth]{aa} 

\usepackage{caption}
\usepackage{subcaption}
\usepackage{graphicx}
\usepackage[varg]{txfonts}
\usepackage{natbib}
\usepackage{hyperref}
\usepackage[utf8]{inputenc}
\usepackage{xcolor,listings}
\usepackage{lscape}
\usepackage[T1]{fontenc}     
\usepackage{babel}           
\usepackage{orcidlink}
\usepackage{multicol}
\usepackage{threeparttable}

\usepackage{lineno} 
\renewcommand\makeLineNumber{}
\nolinenumbers

\usepackage{hyperref}
\hypersetup{
  colorlinks,
  filecolor=blue,
  citecolor=blue,
  linkcolor=blue,
  urlcolor=blue}

\defcitealias{Watkins22}{W2022}
\defcitealias{Munoz-Mateos15}{MM2015}

\newcommand{\software}[1]{\texttt{#1}}

\newcommand*\mean[1]{\overline{#1}}
\newcommand{\csgcolor}{grey\xspace}
\newcommand{\sgcolor}{blue\xspace}
\newcommand{\sgetgcolor}{down-pointing\xspace}
\newcommand{\etgcolor}{orange\xspace}
\newcommand{\ltgcolor}{purple\xspace}

\newcommand{\sgmk}{triangles\xspace}
\newcommand{\sgetgmk}{red~triangles\xspace}
\newcommand{\etgmk}{stars\xspace}
\newcommand{\ltgmk}{circles\xspace}

\begin{document}

   \title{The Complete \textit{Spitzer} Survey of Stellar Structure in Galaxies (CS$^4$G)\thanks{The authors of this paper seek to express deep gratitude for the discussions and expertise from the late \textit{Dr. Tom Jarrett}, a pioneering infrared astronomer who made invaluable contributions to 2MASS, S$^4$G, WISE and other infrared studies. \\}}

   \titlerunning{CS$^4$G, Disc Galaxies extension.}


    \author{P.~M.~Sánchez-Alarcón \inst{1,2} \fnmsep \thanks{\email{pmsa.astro@gmail.com}}  \orcidlink{0000-0002-6278-9233}
        \and 
        H.~Salo \inst{3} \and J.~H.~Knapen \inst{1,2} \and S.~Comerón \inst{2,1}
        \and J.~Román \inst{1,2} \and A.~E.~Watkins \inst{4} \and R.~J.~Buta \inst{5}
        \and S.~Laine \inst{6} \and J.~M.~Falcón-Ramírez \inst{2,1}
        \and M.~Anetjärvi \inst{3,7}
        \and E.~Athanassoula \inst{8}  \and A.~Bosma \inst{8}
        \and D.~A.~Gadotti \inst{9}
        \and J.~L.~Hinz \inst{10} \and  L.~C.~Ho \inst{11,12} \and B.~W.~Holwerda \inst{13} 
        \and J.~Janz \inst{14,3,15} \and T.~Kim \inst{16} \and J.~Koda \inst{17} \and J.~Laine \inst{3,18} \and  E.~Laurikainen \inst{3} 
        \and B.~F.~Madore \inst{19,20} \and K.~Menéndez-Delmestre \inst{21}
        \and R.~F.~Peletier \inst{22}  \and M.~Querejeta \inst{23} 
        \and A.~Ruokanen \inst{3}  \and K.~Sheth \inst{24} \and D.~Zaritsky \inst{25} 
        }

    \institute{Instituto de Astrofísica de Canarias, c/ Vía Láctea s/n, E-38205, La Laguna, Tenerife, Spain
    \and 
        Departamento de Astrofísica, Universidad de La Laguna, E-38206, La Laguna, Tenerife, Spain
    \and 
        Space Physics and Astronomy Research Unit, University of Oulu, Pentti Kaiteran katu 1, FI-90014 Oulu, Finland
    \and 
        Centre of Astrophysics Research, School of Physics, Astronomy and Mathematics, University of Hertfordshire, Hatfield, UK
    \and 
        Department of Physics and Astronomy, University of Alabama, Box 870324, Tuscaloosa, AL 35487, USA    
    \and 
        IPAC, Mail Code 314-6, Caltech, 1200 E. California Blvd., Pasadena, CA 91125, USA
    \and 
        Max-Planck-Institut für Extraterrestriche Physik, Giessenbach-Str. 1, D-85748 Garching, Germany
    \and 
        Aix Marseille Univ, CNRS, CNES, LAM, Marseille, France
    \and 
        Centre for Extragalactic Astronomy, Department of Physics, Durham University, South Road, Durham DH1 3LE, UK
    \and 
        MMT Observatory, University of Arizona, 933 N Cherry Ave, Tucson AZ 85721
    \and 
        Kavli Institute for Astronomy and Astrophysics, Peking University, 12 Beijing 100871, P. R. China
    \and 
        Department of Astronomy, School of Physics, Peking University, Beijing 100871, P. R. China
    \and 
        University of Louisville, Department of Physics and Astronomy, 102 Natural Science Building, 40292, KY Louisville, USA
    \and 
        Finnish Centre of Astronomy with ESO (FINCA), Vesilinnantie 5, FI-20014 University of Turku, Finland
    \and 
        Specim, Spectral Imaging Ltd., Elektroniikkatie 13, FI-90590 Oulu, Finland
    \and 
        Department of Astronomy and Atmospheric Sciences, Kyungpook National University, Daegu 702-701, Republic of Korea
    \and 
        Department of Physics and Astronomy, Stony Brook University, Stony Brook, NY 11794-3800, USA
    \and 
        Normet Oy, Elektroniikkatie 8, FI-90590 Oulu, Finland
    \and 
        The Observatories, Carnegie Institution for Science, 813 Santa Barbara Street, Pasadena, CA 91101, USA
    \and 
        Department of Astronomy \& Astrophysics, University of Chicago, 5640 South Ellis Avenue, Chicago, IL 60637, USA
    \and 
        Observatório do Valongo, Universidade Federal do Rio de Janeiro, Rio de Janeiro, Brazil
    \and 
        Kapteyn Astronomical Institute, University of Groningen, PO Box 800, NL-9700 AV Groningen, the Netherlands
    \and 
        Observatorio Astronómico Nacional (IGN), C/Alfonso XII, 3, E28014 Madrid, Spain
    \and 
        NASA Headquarters Mary W. Jackson Building, 300 E Street SW, Washington, DC 20546, USA
    \and 
        Steward Observatory and Department of Astronomy, University of Arizona, 933 N. Cherry Ave., Tucson, AZ 85721, USA    
    }

    \authorrunning{Sánchez-Alarcón,~P.~M. et al.}

   \date{Received 7 August 2024 / Accepted 28 February 2025}

 
  \abstract
   {The  \textit{Spitzer} Survey of Stellar Structure in Galaxies (S$^4$G), together with its Early Type Galaxy (ETG) extension, stand as the most extensive dataset of deep, uniform mid-infrared (mid-IR; 3.6 and 4.5$\,\mu$m) imaging for a sample of $2817$ nearby ($d<40 \,$Mpc) galaxies. However, the velocity criterion used to select the original sample results in an additional 422 galaxies without H\,{\sc i} detection that ought to have been included in the S$^4$G on the basis of their optical recession velocities.}
   {In order to create a complete magnitude-, size- and volume-limited sample of nearby galaxies we collect $3.6\,\mu$m and $i$-band images using archival data from different surveys and complement it with new observations for the missing galaxies. Since most, but not all, of these galaxies have a Hubble type in Hyperleda $T_\mathrm{HL}>0$, we denote the sample of these additional galaxies as Disc Galaxy (DG) extension. We present the Complete \textit{Spitzer} Survey of Stellar Structure in Galaxies (CS$^4$G), encompassing a sample of $3239$ galaxies (S$^4$G$+$ETG$+$DG) with consistent imaging, surface brightness profiles, photometric parameters, and revised morphological classification.}
   {Following the original strategy of the S$^4$G survey, we produce masks, surface brightness profiles, and curves of growth using masked $3.6\,\mu$m and $i$-band images. From these profiles, we derive the integrated quantities: total magnitude, stellar mass, concentration parameter, and galaxy size, converting between optical \textit{i}-band and $3.6\,\mu$m. We re-measure these parameters also for the S$^4$G and ETG to create a homogenous sample. We present new morphological revised $T$-types and we showcase mid-IR scaling relations for the stellar mass, galaxy size, concentration index, and morphological type.}
   {Our new masking procedure increases the number of pixels masked out by a factor of five, improving the masking of fainter regions over previous S$^4$G data. Our photometric parameters from \textit{i}-band imaging yield measurements consistent with the original sample (S$^4$G) and its ETG extension in the $3.6\,\mu $m band. The new DG extension consists of galaxies with a wide morphological range ($-5<T_{\rm HL}<10$) and a mass range of $6<\log(M_*/M_\odot)<11$. The galaxies in the DG  sample have an average mass of $\log(M_*/M_\odot) = 9.21$, an average galaxy isophotal radius at $25.5\,$mag\,arcsec$^{-2}$ of $R_{25.5}=7.1\,$kpc, and an average concentration index of $C_{82} = 2.92$.}
   {We complete the S$^4$G sample by incorporating 422 galaxies into the original dataset. The new galaxies constitute 15\% of the total previous sample (S$^4$G+ETG), but in the lower-mass range ($M_*<10^{9}M_\odot$), the disc galaxy extension increases the sample by 36\%. The CS$^4$G includes at least 99.94\% of the complete sample of nearby galaxies, meeting the original selection criteria based on a comparison with the NED database. We make the images and surface brightness profiles available to the community together with the conjunct catalogue of the whole CS$^4$G dataset with consistent photometric measurements for 3239 galaxies. The CS$^4$G will enable a wide set of investigations into galaxy structure and evolution, and complement optical, near-infrared and mid-IR imaging to be obtained in coming years with \textit{Euclid}, Rubin, \textit{Roman} and other research projects.}

   \keywords{galaxies: evolution - galaxies: photometry - galaxies: disc - galaxies: structure - galaxies: statistics}

   \maketitle
%
\section{Introduction}
\label{sec:Intro}
The local Universe offers an essential sample for studying galaxy formation and evolution. Most local galaxies have substantial angular size \citep[e.g. ][]{Jarret19, Moustakas23} which facilitates detailed decompositions and morphological classifications. This enables a comprehensive exploration of the roles played in galaxy evolution by structural components such as bulges, bars, rings, or lenses. 

The \textit{Spitzer} Survey of Stellar Structure in Galaxies \citep[S$^4$G;][]{Seth10} is the most extensive dataset of deep, uniform mid-infrared (mid-IR) imaging for a sample restricted by volume, magnitude, and size ($d<40\,\mathrm{Mpc}$, $|b|>30^{\circ}, m_{B,\text { corr }}<15.5$, and $D_{25}>1^{\prime}$). The S$^4$G is a survey of 2352 galaxies observed in the mid-IR with the $3.6\,\mu$m and $4.5\,\mu$m channels of the infrared array camera \citep[IRAC;][]{IRAC} of the \textit{Spitzer} Space Telescope \citep{Spitzer}. IRAC was an infrared camera that produced simultaneously $5\farcm2\times 5\farcm2$ images in two bands, either in 3.6 and 4.5 $\mu$m or  5.8 and 8.0 $\mu$m, with a pixel scale of $1\farcs2$. After the cryogenic phase, during the warm mission of \textit{Spitzer}, only channels 1 ($3.6\,\mu$m) and 2 ($4.5\,\mu$m) continued to be operational because they were less affected by the higher operating temperature. The $3.6\,\mu$m and $4.5\,\mu$m wavelengths are much less affected by dust than optical bands and thus galaxies imaged in those bands have a nearly constant mass-to-light ratio ($\mathcal{M}/L$) that only varies slightly with age and metallicity. As a result, the 3.6 and $4.5\,\mu$m  bands are ideal for measuring stellar masses of galaxies \citep[see][]{Meidt14, Querejeta15}. Studying these galaxies has contributed to our comprehension of the impact of gas on bar formation \citep{Comeron18, Simon19}, the mechanisms behind disk breaks and truncations \citep{LaineJ14-Breaks, Laine16, Comeron12}, thin and thick disks \citep{Comeron11, Comeron18}, the nature of resonance rings \citep{Comeron14-ARRAKIS}, morphological features \citep{LaineS14-Outskirts, Buta15} and the influence of the environment on the gas content of disk galaxies \citep{LaineJ14-Breaks, Watkins19}. The S$^4$G  plays a crucial role in studying the evolution of galaxies, having led to the publication of numerous research articles over the years \citep[see][hereafter, \citetalias{Watkins22}, for a review]{Watkins22}. Recent studies \citep[e.g, ][]{2023MNRAS.518.2300C, 2024MNRAS.52711777M} emphasise the ongoing relevance of the S$^4$G and reinforce its enduring impact.

However, the original S$^4$G sample suffered from a bias due to the velocity-based selection criterion for distance. The selection was based on H\,{\sc i} measurements, which resulted in excluding gas-poor galaxies. \cite{Seth13-Spitzer} overcame this bias using optical-band spectroscopic redshifts for early-type galaxies (ETGs) with HyperLeda morphological types $T_\mathrm{HL}\leq0$, incorporating 465 new \textit{Spitzer} observations for these missing galaxies into the survey \citepalias[published by][]{Watkins22}. The original survey and its extension (S$^4$G+ETG) contain 2817 galaxies. Nonetheless, the $T$-type criterion for the ETGs ($T_\mathrm{HL}\leq0$) does not include 391 late-type, 20 lenticular, and 11 early-type (based on HyperLeda types) galaxies that lack radio-based velocities but meet the distance criterion using optical velocities, as well as the other S$^4$G selection criteria. These new 422 missed galaxies form the disc galaxy (DG) extension described here. 

To complete the survey with these 422 missing galaxies, we collected archival IRAC images and optical \textit{i}-band images from several surveys (see Sect.~\ref{sec:LTG-sample}). We complement these data with new observations for $18$ galaxies lacking archival images. We follow the original strategy of the S$^4$G survey by conducting the same photometric analysis applied to the previous extension of 465 ETGs by \citetalias{Watkins22}. We apply the methods described in the S$^4$G Pipelines 2 and 3 \citep[P2 and P3; see][hereafter \citetalias{Munoz-Mateos15}]{Munoz-Mateos15}. This analysis includes masking images, producing intensity and surface brightness profiles, obtaining position angle and ellipticity profiles, and computing total magnitudes, stellar masses, measures of central concentration, and measures of galaxy size. \citetalias{Munoz-Mateos15} extensively outlined the applications of this structural analysis, all of which are still relevant to this extended sample of gas-poor galaxies that includes mostly disk and dwarf galaxies. 

In the current landscape of extensive deep surveys, notably focused on exploring vast cosmic volumes \citep[e.g., the Hyper-Suprime Cam Subaru Strategic Program, the Vera C. Rubin Observatory’s Legacy Survey of Space and Time, the \textit{Roman} Space Telescope, and the \textit{Euclid} Mission][]{HSC-SSP,LSST,Roman,Euclid}, a complete census of late-type galaxies at redshift $z=0$ becomes crucial for comparisons between young and present-day galaxies. 
\renewcommand{\thefootnote}{\alph{footnote}}

In this work, we present the Complete \textit{Spitzer} Survey of Stellar Structure in Galaxies (CS$^4$G) by adding an extension of 422 galaxies which completes the final sample of 3239 galaxies. This complete dataset serves as the basis for future analyses, including a detailed analysis of morphology and multi-component decompositions. In Sect.~\ref{sec:LTG-sample} we give an overview of the surveys and instruments used for the acquisition of the IR and $i$-band images. In Sect.~\ref{sec:CS4G} we describe the completeness of the CS$^4$G sample. In Sect.~\ref{sec:preparation} we review the methods used for the photometric analysis, the creation of masks, and the sky background estimation. In Sect.~\ref{sec:IR-conversion}, we present the empirical photometric conversion between $i$-band and 3.6$\,\mu$m. In Sect.~\ref{sec:P3}, we explain the radial and integrated photometric parameters derived from the images. In Sect.~\ref{sec:Scaling} we explore the scaling relationships among the photometric parameters within this new sample. We list the conclusions of our results in Sect.~\ref{sec:conclusions}. Finally, in Sect.~\ref{sec:DR} we explain the data products released with this work. Throughout, we use the AB photometric system and we assume the $\Lambda$CDM model with a Hubble-Lemaître constant $H_0=75\footnote{We choose this value to be consistent with \cite{Seth10}.}\,$km\,s$^{-1}\,$Mpc$^{-1}$, and a matter density parameter $\Omega_{\mathrm{m}}=0.3$. Morphological types used throughout the paper are either HyperLeda morphological types, denoted as $T_\mathrm{HL}$, or revised $T$-types presented in Sect.~\ref{sec:P3}, denoted as $T$, which are in the Comprehensive de Vaucouleurs revised Hubble-Sandage (CVRHS) system \citep{Buta15}. 
\renewcommand{\thefootnote}{\arabic{footnote}}

\begin{table}[t!]
    \centering
    \caption{Survey and instrument properties.}
    \begin{threeparttable}
    \begin{tabular}{p{1.8cm}p{1.2cm}p{1.2cm}p{1.2cm}p{1.2cm}}
        \hline \hline \\[-8pt]
        Survey & Galaxies & Scale & FWHM & Depth   \\
        \hline \hline \\[-8pt]
        IRAC & 44 & $0\farcs600$ & $2\farcs1$ & 26.8\\
        IRAC\tablefootmark{a} & 11 & $0\farcs750$ & $2\farcs1$ & 26.8\\
        DES & 102 & $0\farcs263$ & $0\farcs9$ & 28.4 \\
        LS & 169 & $0\farcs263$\tablefootmark{b} & $1\farcs0$  & 27.9\\
        SDSS & 77 & $0\farcs396$ & $1\farcs1$ & 26.4\\
        LT & 15 & $0\farcs150$ & $1\farcs4$   & 28.3\\
        NTT & 3 & $0\farcs241$ & $1\farcs2$& 27.6 \\
        \textit{HST}  & 1 & $0\farcs050$ & $0\farcs1$& 27.5 \\
        \hline
    \end{tabular}
    \tablefoot{The depth column represents the surface brightness limit [$3\sigma,\,10^{\prime\prime}\times10^{\prime\prime}$] in units of mag\,arcsec$^{-2}$. \\
    \tablefoottext{a}{For this set of galaxies, the IRAC images are from S$^4$G fields.}
    \tablefoottext{b}{For 41 galaxies, the mosaics have a pixel scale of $0\farcs269$.}}
    \end{threeparttable}
    \label{tab:survey_sample}
    
\end{table}
\begin{figure*}[h]
    \centering 
    \includegraphics[width=1.0\textwidth]{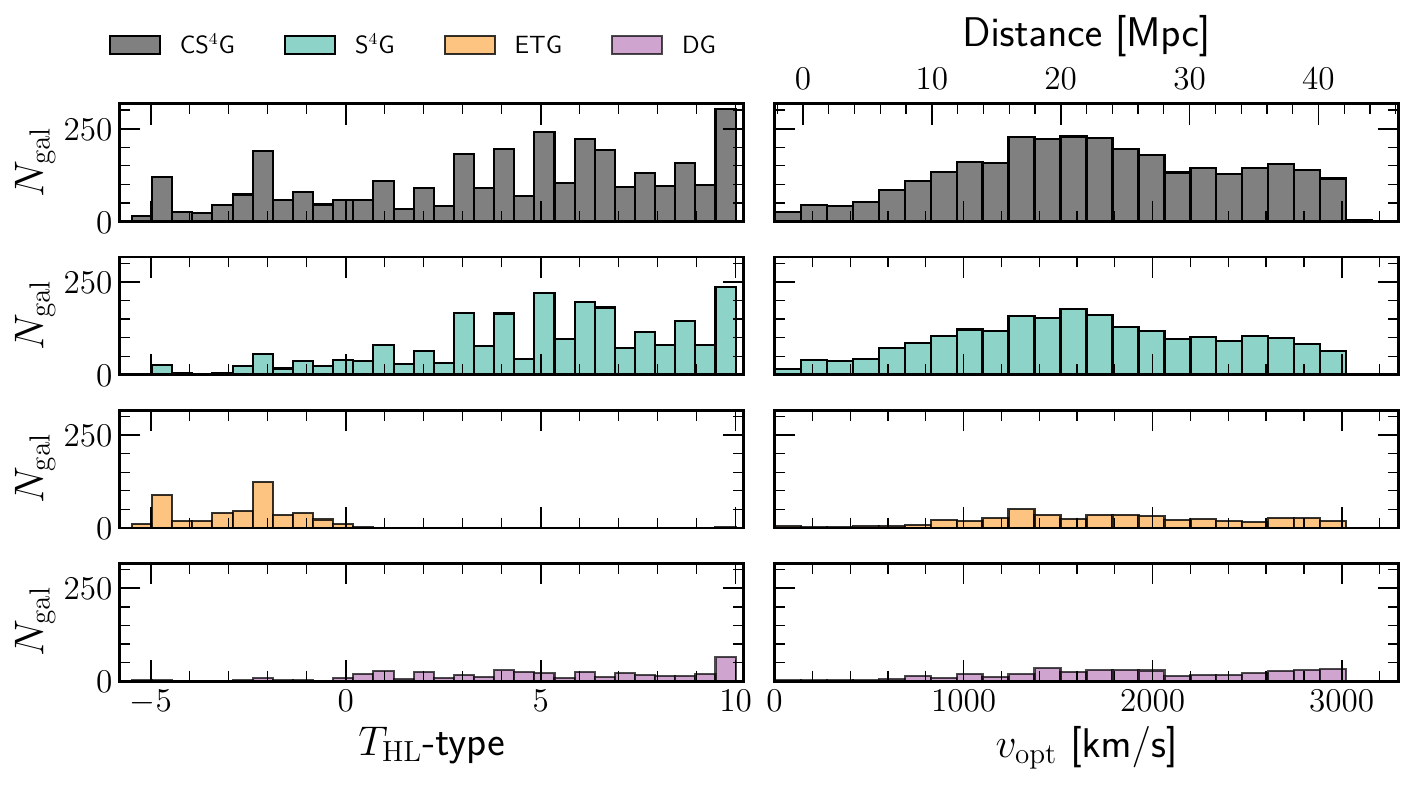}
    \caption{HyperLeda morphological type ($T_{\rm{HL}}$, left) and optical velocity distribution (right) for all the samples. These values are queried from HyperLeda. From top to bottom, each row shows the distribution of the CS$^4$G (\csgcolor), original S$^4$G (\sgcolor), ETG (\etgcolor), and DG (\ltgcolor) samples, respectively.}
    \label{fig:Histogram-Samples}
\end{figure*}

\section{S$^4$G disc galaxy extension sample, observations, and origin of archival data}
\label{sec:LTG-sample}
The S$^4$G disc galaxy (DG) extension sample, as well as the ETG one  \citepalias{Watkins22}, adhere to the same selection criteria as the main S$^4$G project. These criteria include galaxies listed in the HyperLeda\footnote{\url{http://leda.univ-lyon1.fr/}} database \citep{HyperLeda-III} with radial velocities  $v<3000\,$km/s, total extinction-corrected blue magnitudes $m_{B,\rm corr}<15.5$, blue isophotal diameters $1^\prime<D_{25}<30^\prime$, and Galactic latitude $|b|>30^{\circ}$. Thus, the CS$^4$G sample excludes nearby galaxies with angular diameters larger than 30$^\prime$ (LMC, SMC, M33, and UMi Dwarf), and the largest galaxy included is NGC~0055 with a diameter of $29\farcm8$.

The original sample exclusively relied on H\,{\sc i}-derived velocities \citep{Seth10}, which introduced a bias toward gas-rich late~type galaxies. To rectify this bias in the sample, the ETG extension \citep{Seth13-Spitzer, Watkins22} incorporates galaxies with morphological types $T_\mathrm{HL}\leq0$ and optical-band spectroscopic velocities. Nevertheless, this limitation to early types still left out many galaxies (see Appendix \ref{app:sample} for more details): a new search on the HyperLeda database in 2018 led to a list of 422 galaxies with visual-band spectroscopic velocities and meeting the original S$^4$G criteria, forming our DG sample (see Appendix~\ref{app:exclusion} for comments on the exclusion of some galaxies). Of these, 391 galaxies have morphological types $T_\mathrm{HL}\geq0$, 20 galaxies with $-3\leq T_\mathrm{HL}<0$, and 11 galaxies with $T_\mathrm{HL}<-3$. The DG sample consists of these 422 galaxies selected from the HyperLeda database that lacked H\,{\sc i} derived velocities. 

To keep consistency with the original survey, we collect $3.6\,\mu$m and $4.5\,\mu$m band imaging from the \textit{Spitzer} Heritage Archive, maintaining the original photometric bands of the survey. When these bands are not available, we use \textit{i}-band imaging from different surveys. Despite its $3.4\,\mu$m band being similar to IRAC $3.6\,\mu$m, we decided not to use WISE due to its worse angular resolution. We find 55 mid-IR images from the \textit{Spitzer} Heritage Archive, of which 11 are from S$^4$G and ETG footprints, and 367 \textit{i}-band images. For \textit{i}-band imaging, we use the Dark Energy Survey DR2 \citep{DES-DR2} for 102 galaxies, the DESI Legacy Imaging Surveys DR10 \footnote{\url{https://legacysurvey.org/dr10}} \citep[hereafter Legacy Surveys, ][]{DESILegacy} for 169 galaxies, the SDSS DR12 \citep[][]{SDSS, SDSS-III-DR12} for 77 galaxies and archival data from the \textit{Hubble} Space Telescope for one galaxy. For the remaining 18 galaxies we obtain $i$-band images with the Liverpool Telescope (LT) for 15 Northern Hemisphere galaxies, and with the New Technology Telescope (NTT) for three galaxies in the Southern Hemisphere.

We summarise the characteristics of the surveys and instruments in Table~\ref{tab:survey_sample}. We show the number of images of each survey used in this extension, the pixel scale, the average full width at half maximum (FWHM) spatial resolution, and the average surface brightness depth [$3\sigma,\,10^{\prime\prime}\times10^{\prime\prime}$], measured as explained by \cite{Roman20}. In Fig.~\ref{fig:Histogram-Samples} we show the morphological distribution (left column using Hyperleda types $T_{\rm HL}$) and the distance distribution (right column) for the whole CS$^4$G (\csgcolor), the original S$^4$G (\sgcolor), the ETG (\etgcolor), and the DG (\ltgcolor) samples.


\subsection{\textit{Spitzer} Heritage Archive}
We select sample galaxies observed with the \textit{Spitzer} IRAC camera from the \textit{Spitzer} Heritage Archive, either from the cryogenic or the warm phase of the mission. We found 55 galaxies with $3.6\,\mu$m and $4.5\,\mu$m IRAC observations. Of these 55 galaxies, 11 were observed in fields of the original S$^4$G survey. These images have the same properties as those in the S$^4$G survey \citep{Seth10}, with $0\farcs75$/pixel and similar depth. The remaining 44 had observations from different proposals, or mosaics from the \textit{Spitzer} Enhance Image Products with a pixel scale of $0\farcs60$/pixel. The proposal IDs and the Region ID of the frames and mosaics can be found in the headers of the images. Two galaxies, NGC~4516 and NGC~4431, have only $4.5\,\mu$m observations available.

\subsection{Dark Energy Survey}
The Dark Energy Survey \citep{DES} is a Southern Hemisphere imaging survey that aims to probe the origin of the accelerating Universe and help uncover the nature of dark energy. The survey is based on optical and mid-infrared imaging with the Dark Energy Camera \citep[DECam;][]{DECAM} mounted on the 4\,m Blanco telescope at Cerro Tololo Inter-American Observatory in Chile, covering $~5000\,$deg$^2$. We identify 102 galaxies of the DG sample with $i$-band images in the second data release of DES \citep[DR2;][]{DES-DR2}. The DECam has a pixel scale of $0\farcs263$, and the survey depth in the $i$-band is $23.8$ mag (for a $1\farcs95$ diameter aperture at signal-to-noise ratio SNR of 10) with a mean FWHM of $0\farcs88$. 
We access the images via the \texttt{DESaccess}\footnote{\url{https://des.ncsa.illinois.edu/desaccess/}} web application. We use the image cutout service to download images for 102 galaxies, centred on the galaxy, with size $12\times12$\,arcmin$^2$. 

\subsection{DESI Legacy Imaging Surveys}
The DESI Legacy Imaging Surveys \citep{DESILegacy} are a combination of three public projects (the Dark Energy Camera Legacy Survey, the Beijing--Arizona Sky Survey, and the Mayall \textit{z}-band Legacy Survey) that jointly cover $\sim$20000\,deg$^2$ in its latest data release (DR10). This release incorporates additional DECam data from NOIRLab mostly from DES \citep{DES-DR2}, the DELVE Survey\footnote{\url{https://datalab.noirlab.edu/delve}} \citep{DELVE}, and the DECam eROSITA Survey (DeROSITA; PI Dr. Alfredo Zenteno\footnote{\url{http://astro.userena.cl/derositas}}). There are 169 S$^4$G-DG galaxies with $i$-band imaging in the DESI Legacy Imaging Surveys DR10 that are not in the DES DR2 survey. We collect images for these 169 galaxies using the Sky Viewer \footnote{\url{https://www.legacysurvey.org/viewer}} URL cutout service, imposing a size of $\sim 11\times11$\,arcmin$^2$ and at the original pixel scale $0\farcs263$. 

\subsection{Sloan Digital Sky Survey}
The Sloan Digital Sky Survey \citep{SDSS} is probably the most widely used Northern Hemisphere imaging and spectroscopic survey. We use the Data Release 12 \citep[DR12;][]{SDSS-III-DR12} Science Archive Server (SAS) to download individual frames and create mosaics using Swarp \citep{Swarp}. Since DR12 no additional imaging has been incorporated into SDSS. We create mosaics centred at the galaxy with a size of $6\times6$\,arcmin$^2$ for 75 galaxies and for the galaxies NGC~4964 and UGC~8736 that have larger angular size, of $18\times18$\,arcmin$^2$.  

\subsection{Liverpool Telescope}
Of our S$^4$G-DG sample, some galaxies lack mid-IR and $i$-band imaging in public surveys. We observed 15 of these galaxies during the nights of July 16, 2018, December 9, 12, and 14, 2018, January 5 and 12, 2019, March 3 and 16, 2019, and April 11, 2019, with the Optical-Infrared \texttt{IO:O} camera \citep{IO:I} at the $2.0$\,m Liverpool Telescope in the Observatorio del Roque de los Muchachos \citep{LT}. The \texttt{IO:O} camera has a $4096\times4112$ pixel$^2$ CCD, with a pixel scale of $0\farcs15$ and a $10\times10$\,arcmin$^2$ field-of-view.

\subsection{New Technology Telescope}
We observed three further missing galaxies in the Southern Hemisphere lacking archival mid-IR and $i$-band imaging using the EFOSC2 instrument \citep{EFOSC2} at ESO's $3.58$\,m New Technology Telescope (NTT) in Chile during the night of August 11th, 2019. The EFOSC2 instrument is equipped with a $2048\times2048$ pixel CCD, and we use the $2\times2$ binning mode resulting in a pixel scale of $0\farcs2408$ and a $4.13\times4.13$\,arcmin$^2$ field-of-view. 

\subsection{Hubble Space Telescope}
There is existing archival data for one remaining galaxy in the Hubble Legacy Archive
 \citep[Proposal ID: 9395, PI: Marcella Carollo; ][]{NGC2082-HST}. NGC~2082 was observed with the Advanced Camera for Surveys \citep[ACS,][]{ACS-I, ACS-II} using the \textit{F814W} (\textit{I}-band) filter with the Wide Field Channel (WFC) camera of the \textit{Hubble} Space Telescope. The pixel scale of the ACS/WFC camera is $0\farcs050$ providing a $3.4\times3.4$ arcmin$^2$ field of view, sufficient to capture the entire galaxy.\\


\section{Complete \textit{Spitzer} Survey of Stellar Structure in Galaxies}
\label{sec:CS4G}

We present the Complete \textit{Spitzer} Survey of Stellar Structure in Galaxies (CS$^4$G) by adding the DG sample studied in this work to the original S$^4$G \citep{Seth10} and the ETG extension \citepalias{Watkins22}. The CS$^4$G includes 3239 galaxies that are selected using the original S$^4$G criteria (see Sect.~\ref{sec:LTG-sample}). 

To assess the completeness of the CS$^4$G sample, we select galaxies mimicking the criterion used in HyperLeda from the NASA/IPAC Extragalactic Database \citep[NED;][]{Ned}. We use the \texttt{Objects with Parameter Constraints} tool from the NED website\footnote{\url{https://ned.ipac.caltech.edu/byparams}} to select a NED comparison sample. However, this tool does not allow for constraints on the galaxy diameter. From the resulting selection queried by magnitude (\textit{B}-band magnitude brighter than 15.5), volume (redshift less than 0.01), type of object (galaxies) and sky area, we query the diameters. We use an isophotal diameter at $25\,$mag$\,$arcsec$^{-2}$ when available, or the median of all diameters available.  We find 3005 galaxies that meet the criteria of the CS$^4$G according to NED. Of these, 295 are not in the CS$^4$G. This list of galaxies includes the LMC, SMC, M33 and UMi Dwarf galaxies that were excluded from the original S$^4$G due to their large angular size. We also find four objects misclassified as galaxies in NED. SIMBAD and visual inspection of the images show clearly that they are globular clusters of the Milky Way (UGC~09792, ESO~118-031, and Sextans C) and an emission region from the LMC (ESO~056-019). 

We cross-matched the remaining 287 galaxies with the HyperLeda database updated to October 1st, 2024. We find that 285 galaxies fail to meet one or more of the S$^4$G criteria (\textit{B}-band magnitude, diameter, or velocity) according to HyperLeda. Two galaxies (IC~3418 and ESO~084-12) do not have any velocity measurement in HyperLeda, but the other parameters, magnitude, diameter and galactic latitude, meet the criteria. According to NED, these two galaxies have redshifts well below the limit. They could potentially be included in the CS$^4$G. However, to be consistent with the original S$^4$G sample selection criteria, these galaxies are not included since they do not have any velocity measurements in HyperLeda. 

\begin{figure}[t]
    \centering
    \includegraphics[width=\linewidth]{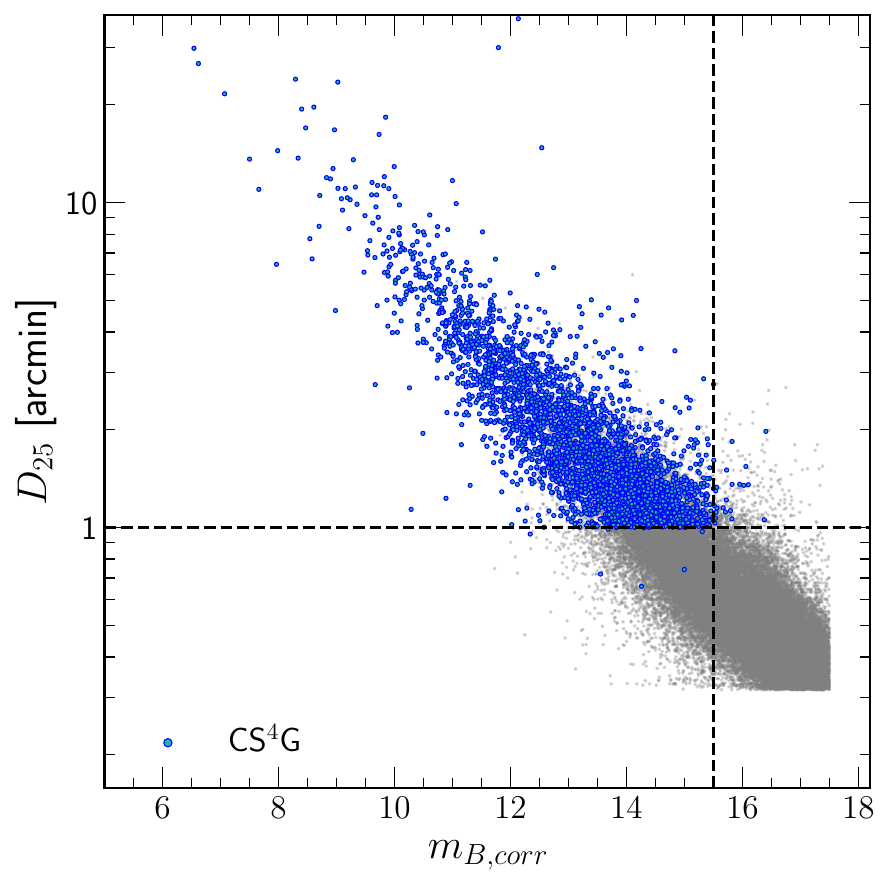}
    \caption{HyperLeda isophotal diameter at 25.5\,mag\,arcsec$^{-2}$ (from \texttt{logd25} parameter) in arcminutes with respect to the \textit{B}-band magnitude (\texttt{btc} parameter). Blue points represent the CS$^4$G sample, and grey dots represent a larger sample from HyperLeda. The vertical and horizontal dashed lines represent the thresholds used to build the CS$^4$G. There are some cases where some galaxies in the CS$^4$G quadrant are not real galaxies meeting the criteria (as explained in Appendix~\ref{app:exclusion})}
    \label{fig:HyperLeda}
\end{figure}

Figure~\ref{fig:HyperLeda} shows the HyperLeda isophotal diameter at 25.5\,mag\,arcsec$^{-2}$  with respect to \textit{B}-band magnitude for the CS$^4$G sample (blue points) and for a larger sample of galaxies extracted from HyperLeda (grey dots). Grey points inside the region of the CS$^4$G are galaxies that fail other criteria while blue points outside the region are galaxies from the original S$^4$G that the newer measurements in HyperLeda made them outside the region. For the sake of completeness, we included them in the CS$^4$G. The CS$^4$G describes the largest and brightest galaxies of the local Universe followed by a smooth transition to fainter galaxies as seen with the representation of a larger sample in grey dots.  

In conclusion, a comparison to NED identifications indicates that there are no new galaxies that meet the original sample selection criteria of the S$^4$G. Thus, the completeness of the CS$^4$G is $100\%$. However, there are two potential galaxies ($0.06\%$ of the sample) that could meet the criteria in the future if the velocity from NED is included and validated in HyperLeda. In this sense, the CS$^4$G contains at least $99.94\%$ of galaxies of the local Universe falling within the CS$^4$G selection criteria. 
 
\begin{figure*}[h]
    \centering
    \includegraphics[width=\textwidth]{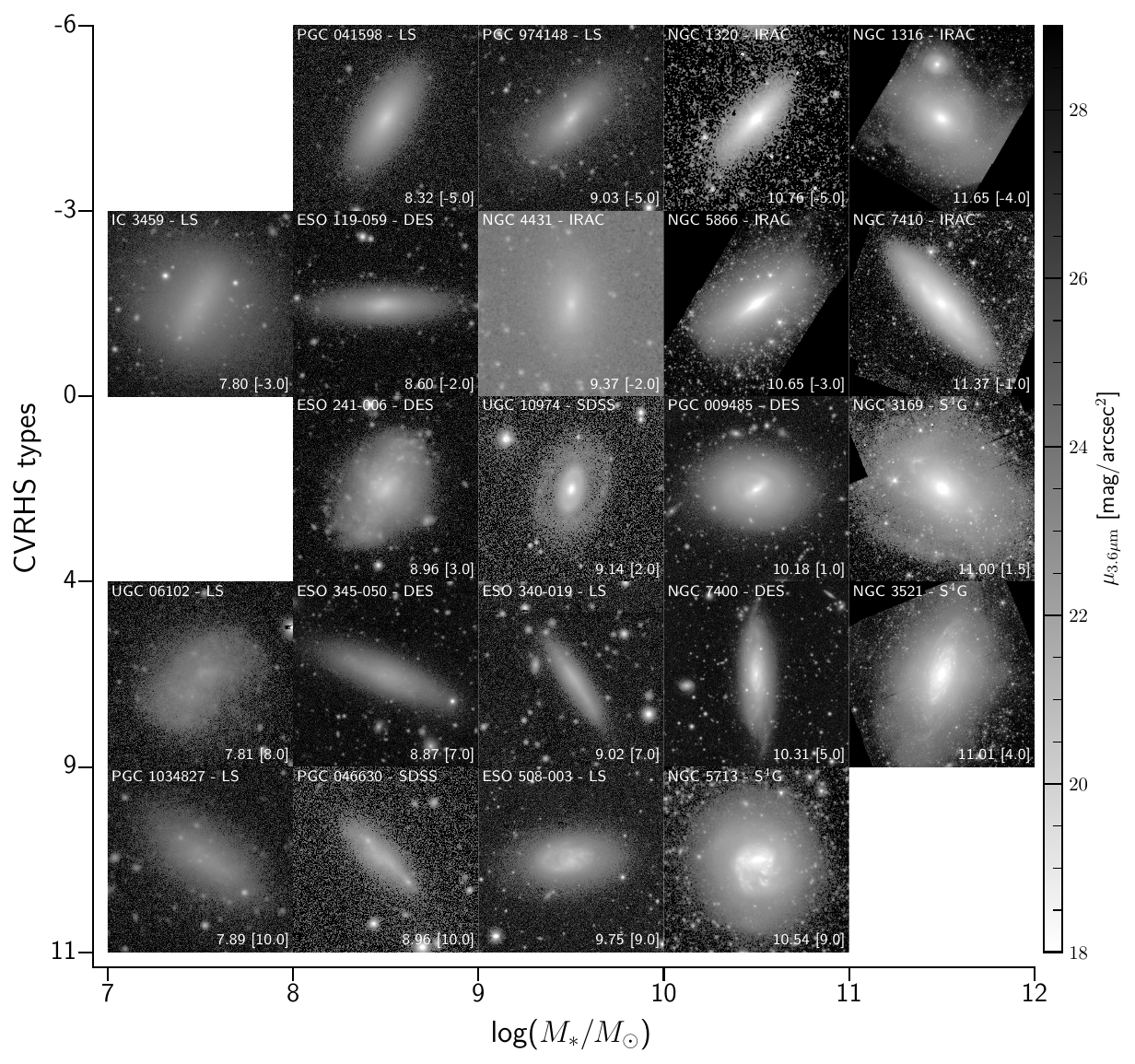}
    \caption{Examples of images of galaxies from the CS$^4$G sample. Galaxies are sorted according to their stellar mass (\textit{x}-axis) and morphological type (\textit{y}-axis) measured as explained in Sect.~\ref{sec:P3}. The galaxy name and the survey or instrument of the image are shown in the upper part of each image. The stellar mass and the revised morphological type (in brackets) are shown in the bottom part of each image.  Each image is centred on the galaxy and the field of view is set to be 1.4 times the isophotal radius at 25.5 mag\,arcsec$^{-2}$ ($R_{25.5}$). The radius, $R_{25.5}$, is measured in the $3.6\,\mu$m band as described in Sect.~\ref{sec:P3}. }
    \label{fig:image_examples}
\end{figure*}

\begin{figure*}[ht!]
    \centering
    \includegraphics[width=\textwidth]{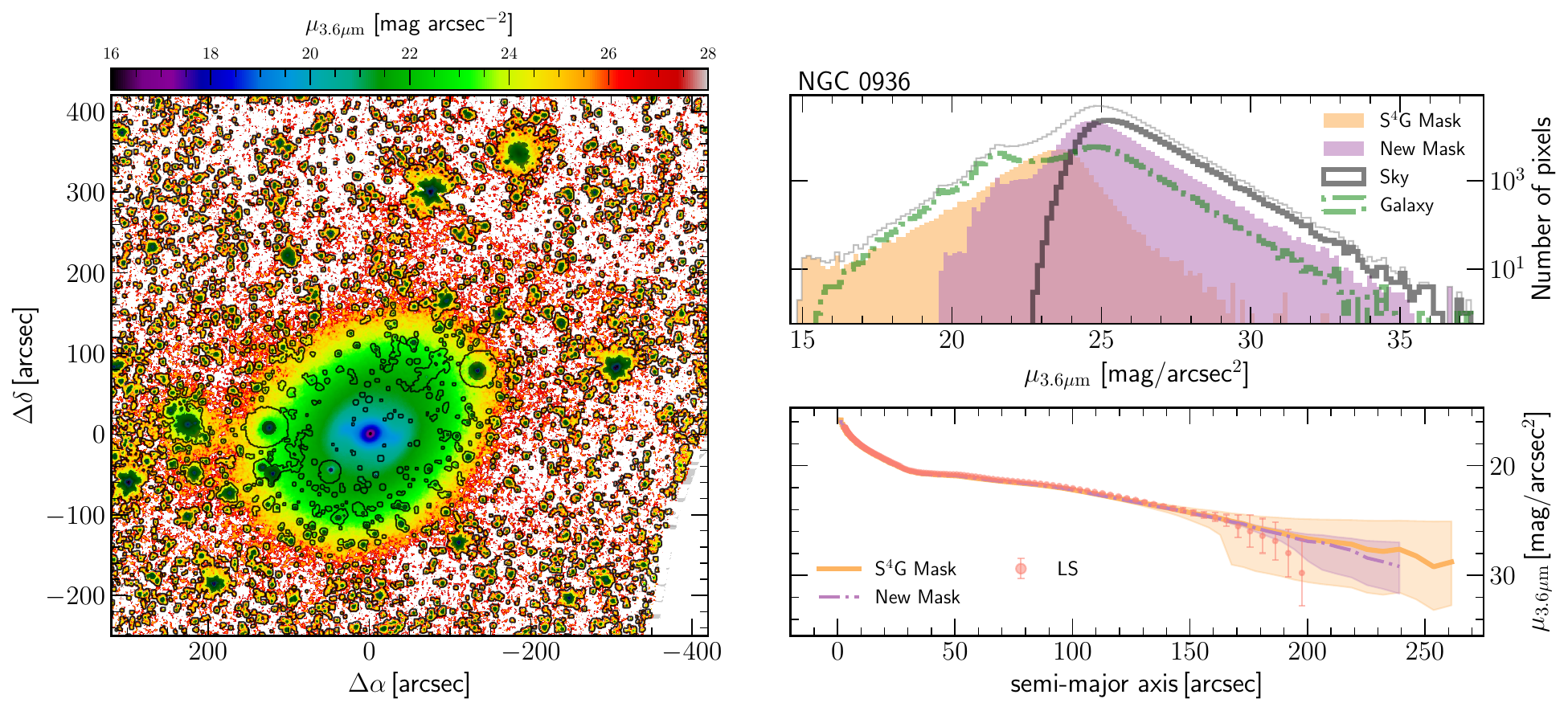}
    \caption{Difference between the masking of the original S$^4$G and our new method. The left panel shows the surface brightness map in the 3.6\,$\mu$m band of the galaxy NGC~0936 from the original S$^4$G image. The shaded region represents the original mask of the S$^4$G, while the black contour lines show the regions masked using the method explained in Sect.~\ref{sec:Mask}. The top right panel shows the histogram of the image (thin grey line) in surface brightness and the different contributions of the pixels in the original S$^4$G mask (orange), the pixels added to the mask with the new method (purple), the pixels within the galaxy (green dotted line) and the sky pixels (black line), which are the pixels that are neither masked nor used to create the profile of the galaxy. The bottom right panel shows the surface brightness profiles in mag\,arcsec$^{-2}$ using the original mask (orange line) and the new mask (purple dashed line). We also show the converted 3.6\,$\mu$m profile (using the recipe from Sect.~\ref{sec:IR-conversion}) using the Legacy Surveys (LS) imaging. The filled region represents the uncertainties due to the sky background subtraction.}
    \label{fig:mask_comparison}
\end{figure*}
\label{sec:Sample}
\section{Data preparation}
\label{sec:preparation}

The original S$^4$G sample and the ETG extension were observed with the \textit{Spitzer} IRAC camera and thus have similar properties, such as the pixel scale ($0\farcs75$), units of MJy\,sr$^{-1}$ and depth ($\mu_{\rm AB,lim}\sim27\,{\rm mag\,arcsec}^{-2}$), as described in \citetalias{Munoz-Mateos15} and \citetalias{Watkins22}. Since the DG extension was observed with a variety of instruments, it does not share these technical details. We adapted Pipeline 1 and 2 \citepalias[P1 and P2, see][]{Munoz-Mateos15} to produce image mosaics for new observations and mask images for the 422 galaxy images in the disc galaxy sample. We follow the procedures used for the original S$^4$G \citep[\citetalias{Munoz-Mateos15},][]{ Salo15} and the ETG extension \citepalias{Watkins22} to have consistent measurements throughout the surveys. We explain in detail the differences in the methods used in the following sections while summarising the similarities to the previous analyses.

\subsection{Pipeline 1. Mosaics}
For the new LT and NTT observations we follow the procedures in Pipeline 1 as described in \citetalias{Munoz-Mateos15} to achieve consistency with the whole survey.  The pipeline initially reduces all the frames, subtracting the dark and bias and correcting the flat field. Then, it aligns the background levels of individual exposures by utilising overlapping regions. Subsequently, it combines all frames following standard dither/drizzle procedures \citep{Drizzle}. The final mid-IR images are delivered in units of MJy\,sr$^{-1}$ and for the optical images in nanomaggies \citep{SDSS-DR1} per pixel (the pixel size can be found in Table~\ref{tab:survey_sample})  with a zero~point of 22.5\,mag\,arcsec$^{-2}$. \\

Figure~\ref{fig:image_examples} shows examples of images of different galaxies from the DG sample, the original S$^4$G and the ETG extension. We use images of the different surveys and instruments in the DG sample to show their differences in angular resolution and depth (see Table~\ref{tab:survey_sample}). We sort the galaxies according to their stellar mass in the \textit{x}-axis and to their morphological type in the \textit{y}-axis (see Sect.~\ref{sec:P3}). The SDSS examples are the images with the noisiest backgrounds, while the DES and LS imaging surveys have more depth and a better angular resolution. 
When selecting galaxies for each panel, we first look for galaxies within the DG sample, and when they are not available, we select them from the original S$^4$G. For panels with multiple options, the galaxy shown is chosen randomly. Blank spaces are regions where there are no galaxies in the CS$^4$G. Despite their differences, the set of images used in the DG samples traces the stellar content of their galaxies similarly to those used in the original S$^4$G.

\subsection{Pipeline 2. Masking}
\label{sec:Mask}
The Pipeline 2 of the S$^4$G and the ETG extension produced masks of contaminating sources, including foreground stars, background galaxies, and scattered light artefacts, using \texttt{SExtractor} \citep[][]{SExtractor} with three different thresholds. Originally, three different masks on the $3.6\,\mu$m mosaic images, with high, medium, and low detection thresholds, were produced to identify sources both far from the target galaxy and overlapping in projection with the galaxy. We adopt a similar strategy to produce masks for the DG sample using a combination of \texttt{SExtractor}(v.2.25.0) masks with different thresholds and configurations to improve the detection of the lower SNR regions (i.e., low surface brightness features).

We first smooth the image using the Bayesian noise reduction technique \software{FABADA} \citep[][]{FABADA} with a standard deviation of the noise measure on the image with hard sigma clipping (rejecting pixels above $2.5\,\sigma$). We then combine three different runs of \software{SExtractor} in the smoothed image. For the first two, we vary the background size (\texttt{BACK\_SIZE}), and detection threshold (\texttt{DETECT\_THRESH}). 
In the first iteration, we run it with an intermediate threshold ($\texttt{DETECT\_THRESH}=1.2$) and large background size ($\texttt{BACK\_SIZE}>50 \times\mathrm{FWHM}$)  which produces good masking of extended regions in the image at intermediate SNR regions. We then run \software{SExtractor} again, now optimised for point source detection, using a lower threshold ($\texttt{DETECT\_THRESH}=0.9$) and smaller background size ($\texttt{BACK\_SIZE}<10 \times\mathrm{FWHM}$). We vary the background size according to the image resolution (measured as FWHM) to increase the detection of smaller point sources in higher-resolution images. 

In the last step, we aim to increase the detection of faint point sources and deblend the sources inside the galaxy. We run \software{SExtractor} on the residual of the image enhanced with \texttt{FABADA} minus a Gaussian smoothed version of the image (with a kernel width of the FWHM). This residual highlights the regions that are brighter than the average of the surrounding area, i.e., the most likely peaks of sources. Then we mask regions compatible with being stars, masking rounder objects weighted by the inverse of the distance to the centre of the galaxy (the closer to the centre of the galaxy, the rounder the region must be to be masked). As a final step, to improve the detection of the fainter regions of the sources, which are more affected by the wings of the point spread function (PSF), we dilate the final mask images using a Gaussian kernel of size $3\times 3\, {\rm pixel}^2$ and $\sigma=1\,$pixel. We visually inspect all the masks produced to confirm that all the external sources are masked. When needed, we edit masks manually to overcome some problematic results, and we make use of colour images when available to disentangle bright regions of the galaxy from other sources, usually higher-redshift sources with redder colours. 

Using the same method described here we compute a new mask image for NGC~0936 from the original S$^4$G. We show its impact on the surface brightness profiles and background characterisation in Fig.~\ref{fig:mask_comparison} where we compare the mask produced by the original technique of the S$^4$G and our new method on the galaxy NGC~0936. The left panel shows the original 3.6\,$\mu$m band image of the galaxy. On top of the image, we show the original S$^4$G mask with darker shaded regions and the new mask computed with our method using contours (black line). The top right panel shows the histogram (thin grey line) of the image but with the pixels converted to surface brightness and with the sky background subtracted. We show the contribution to the histogram of the pixels masked by the original mask image (orange), the pixels added by our new mask (purple), and the pixels of the galaxy (green dashed line). We determined the pixels of the galaxy as those used to derive the radial surface brightness profile of the galaxy (see Sect.~\ref{sec:P3}) up to a distance of $250\,$arcsec where the profile reaches the noise level. The remaining pixels that were not masked and not used to construct the profile of the galaxy constitute the background sky (thick black line). In the bottom right panel, we show the radial surface brightness profile measured using the original S$^4$G mask and the new mask. We also show the radial profile measured using the deepest and widest imaging available, the \textit{i}-band image of the Legacy Survey. We convert the \textit{i}-band profile to the 3.6\,$\mu$m band using our recipe described in Sect.~\ref{sec:IR-conversion}.  

\begin{figure*}[ht!]
    \centering
    \begin{subfigure}[t]{0.48\textwidth}
        \centering
        \includegraphics[width=\textwidth]{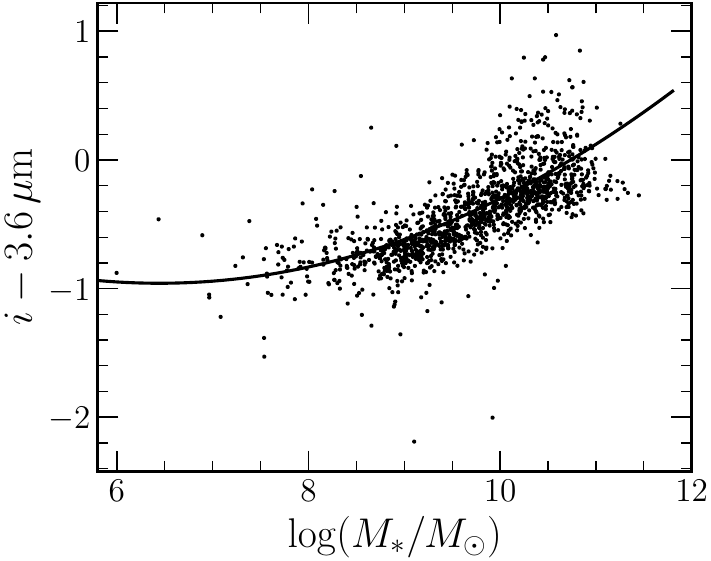}
        \label{fig:mass}
    \end{subfigure}
    \hfill
    \begin{subfigure}[t]{0.48\textwidth}
    \centering
    \includegraphics[width=\textwidth]{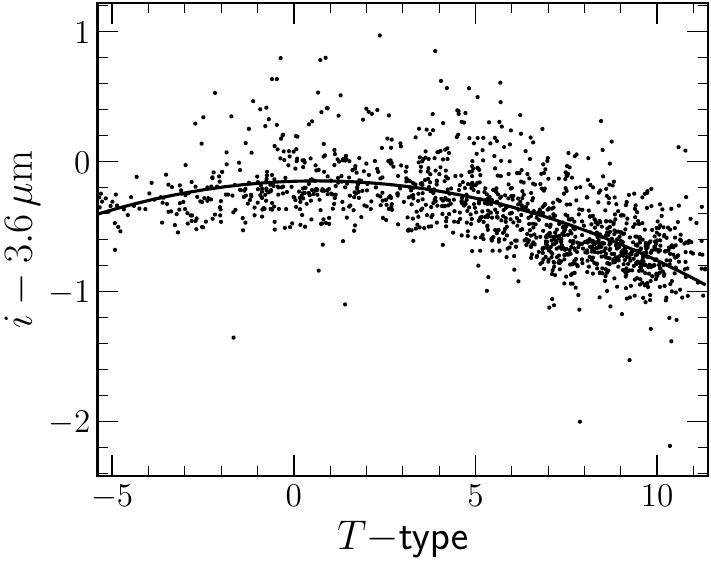}
    \label{fig:type}
    \end{subfigure}
    \caption{Optical to infrared conversion using galaxies in the original S$^4$G sample. The left panel shows the $i-3.6\,\mu{\rm m}$ colour versus the stellar mass relationship. The right panel shows the $i-3.6\,\mu{\rm m}$ colour plotted against the CVRHS revised $T-$types. To have a more homogeneous distribution in the right-hand panel, we added a random offset according to the uncertainties of the morphological revised $T$-types as reported by \cite{Buta15}, from a normal distribution with a width of their uncertainties, $\mathcal{N}(0,0.52)$. }
    \label{fig:combined}
\end{figure*}
Compared with the original S$^4$G, we increase the number of pixels masked by almost a factor of five, from 5\% on the original mask to 24\% in the new mask. The original S$^4$G fails to mask out the faint outskirts of the sources, with an increase of unmasked pixels above $\sim24\,$mag\,arcsec$^{-2}$, where both distributions cross (Fig.~\ref{fig:mask_comparison}). This results in 89\% of pixels with a surface brightness fainter than 22\,mag\,arcsec$^{-2}$ which are not masked out by the original S$^4$G. When measuring the background value, we obtain a difference of 0.4\% between the old and new masks. However, we decrease the error on the background value by 52\% when using the new mask. This is also seen in the profiles panel (Fig.~\ref{fig:mask_comparison}; bottom right), where the error on the profile (the shaded region) is larger for the original S$^4$G profile (orange) than for the new one (purple). The integrated magnitude of the galaxy we measure is $9.54 \pm 0.04$ and $9.55 \pm 0.01\,$mag using the original S$^4$G and new mask, respectively. Both values are compatible, with a small difference of $0.01\,$mag but with an error four times larger when using the original S$^4$G mask. This effect occurs because the boxes used to measure the background value are contaminated with low surface brightness (LSB) pixels from sources not properly masked when using the original S$^4$G mask, which increases the variance of the pixels, increasing the error on the background value. However, not properly masking LSB regions does not significantly affect the measured value of the sky background.  Our new method is more effective in masking the outskirts of sources, i.e. detecting pixels down to a lower SNR (LSB regions) than the original S$^4$G method.

The three profiles shown in the bottom right panel match up to a surface brightness of $26\,$mag\,arcsec$^{-2}$ when the profiles start to diverge. The profiles from IRAC imaging (orange and purple) reach the outskirts of the galaxy and extend down to $\sim29-30\,$mag\,arcsec$^{-2}$, which is expected according to its surface brightness limit. However, they differ due to the mask used to measure the profiles. Using the original S$^4$G mask, we incorporate signal from contaminating sources which results in a brighter tail on the profile. Despite its deeper surface brightness limit (see Table~\ref{tab:survey_sample}), the Legacy Surveys fails to preserve the outskirts of the galaxy. This is an effect of the Legacy Surveys pipeline where the sky background is overestimated and extended objects usually show signatures of destroyed LSB regions. This oversubstraction problem has also been noticed in other works \citep[see][]{2023ApJ...953....7L}. However, this is limited to LSB regions fainter than  $\sim27\,$mag\,arcsec$^{-2}$ and therefore does not affect our results.

\subsection{Pipeline 3. Sky background}
We follow a similar procedure as that in \cite{Salo15}, using a semi-automatic IDL routine to manually select regions that are used to estimate the sky background, typically 20\,--\,30 locations outside the visible galaxy. We avoid placing the boxes on the image edges and in crowded areas. The local sky values in these chosen locations are determined by calculating medians of non-masked pixels within $30\,\rm{pixel}$ by $30\,\rm{pixel}$ boxes. Subsequently, the global sky background (\texttt{sky}) and its uncertainty (\texttt{DSKY}) are derived from the mean and standard deviation of these local values, respectively.



\section{Optical to infrared correction}
\label{sec:IR-conversion}

Since the goal of the DG extension is to provide images that can be used alongside those of the original S$^4$G and the ETG extension, we need to convert from $i$-band to $3.6\,{\rm \mu m}$ AB magnitudes. As a benchmark for the calibration we used the 1388 S$^4$G galaxies that have both an SDSS $i$-band image compiled in \citet{Knapen14} and a well-defined $\mu_{3.6\mu{\rm m}}=26.5\,{\rm mag\,arcsec^{-2}}$ isophote in \citetalias{Munoz-Mateos15}. For each galaxy we derive $i$-band and $3.6\,{\mu{\rm m}}$ asymptotic magnitudes with profiles obtained out to the $\mu_{3.6\mu{\rm m}}=26.5\,{\rm mag\,arcsec^{-2}}$ isophote while keeping the orientation parameters fixed to the values presented in \citetalias{Munoz-Mateos15}. For the $3.6\,\mu{\rm m}$ images, we used the aperture correction described in Eq.~1 of \citetalias{Watkins22}. Pipeline~2 masks were accounted for while producing the profiles.

Figure~\ref{fig:combined} shows the derived $i-3.6\,{\rm \mu m}$ colours as a function of stellar mass (left panel) and morphological $T-$type from \cite{Buta15} (right panel). We fitted the data with order-two polynomials. The resulting relation between the colour and the stellar mass is
\begin{align}
\label{eq:colourmass}
i - 3.6\,{\rm \mu m} =& (0.052\pm0.007) (\log(M_*/M_\odot))^2-\nonumber\\ &-(0.67\pm0.13)\log(M_*/M_\odot)+(1.20\pm0.61),
\end{align}
and between the colour and the $T-$type
\begin{align}
i - 3.6\,{\rm \mu m} = &-(0.0070\pm0.0004)T^2+\nonumber\\&+(0.009\pm0.004)T-(0.152\pm0.011).
\end{align}
The two expressions are represented as continuous black curves in their respective figures. The first expression has a root-mean-square deviation of $0.17\,{\rm mag}$ and the second one of $0.19\,{\rm mag}$.

\begin{figure*}[h]
    \centering
    \includegraphics[width=\textwidth]{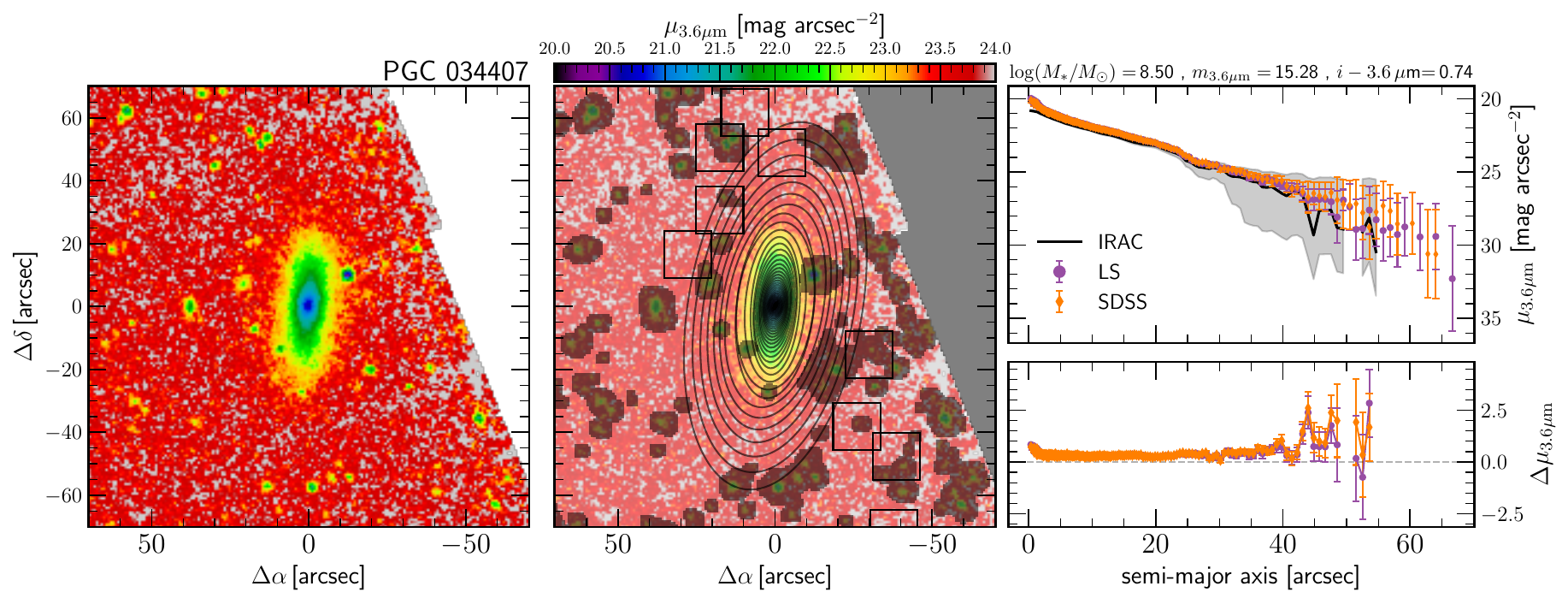}
    \caption{Surface brightness maps and profiles. The left panel shows the surface brightness map of PGC~34407 in the $3.6\,\mu$m band. The middle panel shows the surface brightness map overlaid with the mask image with shaded regions, and the elliptical apertures used to measure the radial profile. The top right panel shows the radial surface brightness profile derived using the $3.6\,\mu$m band (black curve), the \textit{i}-band from the Legacy Imaging Surveys (purple circles), and \textit{i}-band from the SDSS (orange diamonds). The radial profiles of both the \textit{i}-band images have been converted to $3.6\,\mu$m  using the recipe described in Sect.~\ref{sec:IR-conversion}. The lower right panel shows the difference between the mid-IR profile and the converted optical profiles.}
    \label{fig:P3-profile}
\end{figure*}
We convert our \textit{i}-band apparent magnitudes to 3.6 $\mu$m apparent magnitudes using the mass relationship (Eq.~\ref{eq:colourmass}), which has the lowest sum of the squares residual. Since we need the stellar mass of a galaxy to estimate the $i-3.6\,\mu$m colour and the stellar mass is one of the parameters we want to derive from our images, we apply an iterative procedure to estimate the colours. We first assume that the colour $i-3.6\,\mu \rm{m}=0$ and then estimate the stellar mass of the galaxy (see Sect.~\ref{sec:StellarMass}). With this first mass estimate, we calculate the  $i-3.6\,\mu$m colour using Eq.~\ref{eq:colourmass}, we transform the \textit{i}-band magnitudes to $3.6\,\mu$m and calculate the mass with the 3.6$\,\mu$m magnitude again. We repeat this process until we reach convergence in the  $i-3.6\,\mu$m colour. We reach the convergence after 5\,--\,9 iterations to the fifth decimal value. This uncertainty is well below our uncertainty in the integrated magnitudes.


\section{Derived Quantities}
\label{sec:P3}
We follow the original S$^4$G procedures \citep[\citetalias{Munoz-Mateos15}, ][]{Salo15} along with those in the ETG extension \citepalias{Watkins22} to derive photometric quantities. We measure radial profiles of flux and surface brightness, position angle and ellipticity profiles for the DG sample. For the whole set of galaxies in the CS$^4$G, we measure asymptotic magnitudes, concentration indices, and sizes. The parameters from the three samples are homogenised using the same methods. 

All the parameters derived from this analysis can be found in an electronic table accessible from the CDS and IRSA (see Sect.~\ref{sec:DR}).

\subsection{Radial profiles}
To derive radial profiles, we use the implementation of the iterative ellipse-fitting method described by \cite{Jedrzejewski} in the \texttt{Photutils} package \citep[][\texttt{v1.5.0}]{photutils-v1.5.0} of \texttt{Astropy} \citep[][\texttt{v5.1}]{astropy:2013, astropy:2018, astropy:2022}. This method is efficient in tracing the structure of a galaxy, following bars and other features found in the isophotes (e.g. \citealp{Salo15}, \citetalias{Watkins22}, \citealp{Sanchez-Alarcon23}). This is the same method used in the original sample \citepalias{Munoz-Mateos15}, but implemented in the Python programming language. 

We first estimate the galaxy orientation using the image moments on the masked image to improve the success of the fitting procedure, which requires a first ellipse to initialize. Previously, the implementation in \cite{Salo15} of the same method iterated until the fitting was done through the entire galaxy but sometimes it stopped prematurely, missing the outskirts or failing to fit the inner regions, and forcing a restart of the fitting procedure. To accelerate the process and make it more efficient, we estimate the maximum radius of the profile as the radius where we reach the value that was on the order of the sky level. This way, the fitting is more robust and successful for almost the whole sample ($\gtrsim 90\%$). After a visual inspection of the profile, we decide the maximum radius of the profile. We keep the centre position fixed through the whole fitting procedure and measure the position angle (PA) and the ellipticity ($\epsilon$) for each radial bin. We determine the centre using the same method as in \cite{Salo15}, in which a first guess is done manually and the centre is refined as the location where the brightness gradient is zero. We increase the radius logarithmically by $2\%$ for each radial bin.  This adapted step size allows us to have optimal radial resolution in the inner regions, with wider ellipses in the outskirts reaching higher SNR. We estimate the uncertainties as the quadratic sum of the error on the fit procedure, and the uncertainty of the sky value measured in P3.

We show an example of the surface brightness profile measured of PGC~34407 in the 3.6$\,\mu$m band in Fig.~\ref{fig:P3-profile}. The left panel shows the surface brightness map of the 3.6$\,\mu$m image and the middle panel shows the same map with the mask regions shaded and the elliptical apertures used to measure the profiles, along with the boxes used to measure the sky background value and its uncertainty. The top right panel shows the 3.6$\,\mu$m surface brightness radial profiles of the IRAC imaging, and the profiles from the optical LS and SDSS surveys, converted to 3.6\,$\mu$m using our recipe described in Sect.~\ref{sec:IR-conversion}. In the lower right panel, we show the difference between the converted optical profiles and the IRAC profile. We use the same elliptical apertures to measure the three profiles. The elliptical isophotes successfully follow the orientation of the inner and the outer regions of the galaxy and the profiles reach the faint outskirts. We observe that the three profiles exhibit consistent behaviour. We see the expected differences, in the inner region caused by the PSF and in the outer part due to the depth of the different images. We reach surface brightnesses of $\gtrsim 25.1$ and $26.2\,$~mag$\,$arcsec$^{-2}$  [$3\sigma,\,10^{\prime\prime}\times10^{\prime\prime}$] in the mid-IR and \textit{i}-band, respectively, for 95\,\% of the galaxies. Some galaxies have contaminating light from brighter neighbouring sources (i.e., bright stars or galaxies) that would require a model subtraction to describe the faintest regions. 

We do not measure radial profiles for the S$^4$G and ETG galaxies, instead, we used the ones derived from \citetalias{Munoz-Mateos15} and \citetalias{Watkins22}. However, when converting profiles to physical units, we use updated distances.

\subsection{Asymptotic magnitudes}
We derive asymptotic magnitudes from the curve of growth (hereafter, c.o.g.) of the 3.6\,$\mu$m AB magnitude following the same procedure as in \citetalias{Munoz-Mateos15} and \citetalias{Watkins22}. We measure the cumulative sum over the flux within elliptical apertures with constant ellipticity~($\epsilon$), and position angle~(PA), and logarithmically grown by 2\% in semi-major axis. We set the orientation values, PA and $\epsilon$,  to match the outskirts of the galaxy which were measured by P3. Given the depth of the DG extension, the c.o.g. flattens in the outskirts of the galaxy and the relationship between the local gradient and the magnitude enclosed within each elliptical aperture becomes roughly linear. We can measure the asymptotic magnitudes as the $y$-intercept of the linear fit of the last points on the magnitude-local gradient plane. For galaxies with \textit{i}-band images, we convert the \textit{i}-band magnitudes to $3.6\,\mu$m magnitudes.

\citetalias{Watkins22} show in their Fig.~5 a representation of the same method as what was used in this work to measure the asymptotic magnitudes for two galaxies in the sample with different inclinations, luminosity, and radial extent. They also discuss the different effects on the uncertainty of this method introduced in the magnitude derived. This method is accurate enough that the uncertainty introduced is smaller than the systematic uncertainty induced by the sky subtraction. 

\subsection{\label{sec:StellarMass}Absolute magnitudes and stellar masses}

To transform apparent magnitudes to absolute magnitudes, we query distances from the NED database \citep{Ned}. We use redshift-independent measurements when available (for 97 galaxies, $\sim 23\%$) and redshifts when not (for 325 galaxies, $\sim 77\%$). There are four galaxies with redshift-independent measurements with very high and unexpected values, of the order of $\sim100\,$Mpc. For these galaxies, we use another distance indicator such as the redshift when available or optical velocities from HyperLeda (see Appendix~\ref{app:distace} for discussion). 
We investigated the use of Cosmic-Flow corrected distances \citep{2015MNRAS.450..317C} and found that, for nearby galaxies, redshift-independent distances or simple redshift-derived estimates provide a more reliable determination of distance.

We also updated the distances of the galaxies in the S$^4$G and ETG samples. We find 1920 (82\%) and 247 (53\%) galaxies with redshift-independent measurements for the S$^4$G and ETG, respectively. This results in 2264 (70\%) galaxies with redshift-independent measurements for the entire CS$^4$G. 

Then, we transform the 3.6\,$\mu$m absolute magnitudes to stellar masses.  As reported by \cite{Querejeta15}, in the 3.6\,$\mu$m band we can assume a constant mass-to-light ratio $\Upsilon_{3.6 \mu \mathrm{m}}$ which varies by 10\%--30\% on integrated galaxy scales. This uncertainty does not affect our general results in terms of statistical behaviour in trends. We assume a Chabrier initial mass function \citep{Chabrier-IMF} and use a constant mass-to-light ratio of  $\Upsilon_{3.6 \mu \mathrm{m}} \sim 0.6\left(\mathcal{M}_{\odot} / L_{\odot}\right)_{3.6 \mu \mathrm{m}}$ \citep[see][]{Meidt14, Querejeta15, Comeron18} to estimate the stellar mass for all the galaxies. 

\subsection{Concentration indices}

The spatial distribution of light contains valuable information about the morphological classification of galaxies and their evolution. Concentration indices can be directly measured from the light distribution of a galaxy. Again following \citetalias{Munoz-Mateos15} and \citetalias{Watkins22}, we measure the concentration parameters $C_{31}$ \citep{deVaucouleurs77} and $C_{82}$ \citep{Kent85}, defined as:
\begin{align}
C_{31}&=R_{75} / R_{25} \mathrm{,\, and}\\
C_{82}&=5 \log \left(R_{80} / R_{20}\right) \text {, }
\end{align}
where $R_x$ is the radius containing $x \%$ of the total luminosity of the galaxy. As in \citetalias{Munoz-Mateos15} and \citetalias{Watkins22}, to avoid assumptions about the shapes of the light profiles, we measure $R_x$ from the c.o.g. We extrapolate the total luminosities of galaxies out to infinity rather than measuring them within set apertures. 

We interpolate the c.o.g. to improve the resolution of the profile and increase the accuracy of the measurement. We estimate the uncertainty of the concentration indices for each galaxy by performing a  Monte~Carlo sampling on the value of the sky, modelling $1000$ random sky values from a normal distribution $\mathcal{N}(\rm{SKY ; DSKY})$, measuring the concentration indices and estimating the error on the value as the standard deviation from each value. 

\subsection{Galaxy sizes}
\label{sec:Size}
We measure the effective (or half-light) radius, $R_{\rm e}$, along with the two isophotal radii, $R_{25.5}$ and $R_{26.5}$, denoting the radii at which the surface brightness profile of the galaxy reaches 25.5 and 26.5 mag\,arcsec$^{-2}$, respectively, in the $3.6\,\mu$m and \textit{i}-bands. We also apply an inclination correction to measure the isophotal radius at 25.5 and 26.5 mag\,arcsec$^{-2}$ ($R_{25.5,\rm corr}$ and $R_{26.5,\rm corr}$) in both bands. We follow the correction used in \citetalias{Watkins22} and multiply the intensity profiles by the axial ratio. These radii are measured from the major axis of the elliptical apertures. We measure $R_{\rm e}$ using the c.o.g. and employing the asymptotic magnitudes to set the total flux of the galaxy. We estimate the uncertainties of the values using the same Monte~Carlo strategy as in the concentration indices. We measure the galaxy sizes in the  $3.6\,\mu$m and  $4.5\,\mu$m channels when available, and for those galaxies with only \textit{i}-band we measure it from the \textit{i}-band profile and the transformed $3.6\,\mu$m profile. 

\subsection{Average surface brightness levels}

Using the fitted radial surface brightness profile, we measure the mean surface brightness within the effective radii ($\mean{\mu}_{\rm e}$) and within 1\,kpc ($\mean{\mu}_{1\,{\rm kpc}}$). We interpolate the profile to have a higher resolution and extract more precise values. The uncertainties are again estimated using the Monte~Carlo strategy. We measure these values for the DG but also for the entire S$^4$G and the ETG extension.


\subsection{Morphological revision}
\label{subsec:Morphology}

The morphological classifications of the sample galaxies have been made in a comprehensive version of the 
\cite{Vaucouleurs59} revised Hubble-Sandage (CVRHS) system \citep{Buta15}. 
This includes the recognition of many features of interest to extragalactic observers that were considered mere details in the past. The hallmark of the system is the prominence of galactic rings and lenses and the ease with which these can be added to the system. Rings and lenses are considered primary tracers of galactic secular evolution \citep{Knapen12}.
The CVRHS classification also includes barlenses \citep{2011MNRAS.418.1452L, 2017A&A...598A..10L} denoting the inner lens-like components, actually forming part of the bar, embedded in thin bars in massive galaxies.
The galaxies were classified using the 3.6\,$\mu$m and \textit{i}-band images presented in this work, together with \textit{g}-band images from the SDSS and LS when available. The \textit{g}-band is closest in wavelength to the \textit{B}-band, the historical band originally used for galaxy morphological study.

A noteworthy finding of our study is that the DG galaxy sample contains a significant number of extreme late-type spirals (i.e., types Scd and later). Typically, such galaxies tend to be fairly rich in H$\,${\sc i} \citep{Buta94}. Similarly, there appears to be an unusual number of spindles (edge-ons) in the sample.

Images in the \textit{g}-band were not available for all of the galaxies, and also some images had poor seeing. Since the classifications are based on only a single examination (i.e., without a second classification of the same objects using the same images), we assume that the uncertainty is $\sigma(T)=0.7$ stage intervals \citep{Buta15,Buta19-AMIGAS}. The resulting classification is shown in Table~\ref{tab:CVRHS}.

\begin{figure*}[t]
    \centering 
    \includegraphics[width=\textwidth]{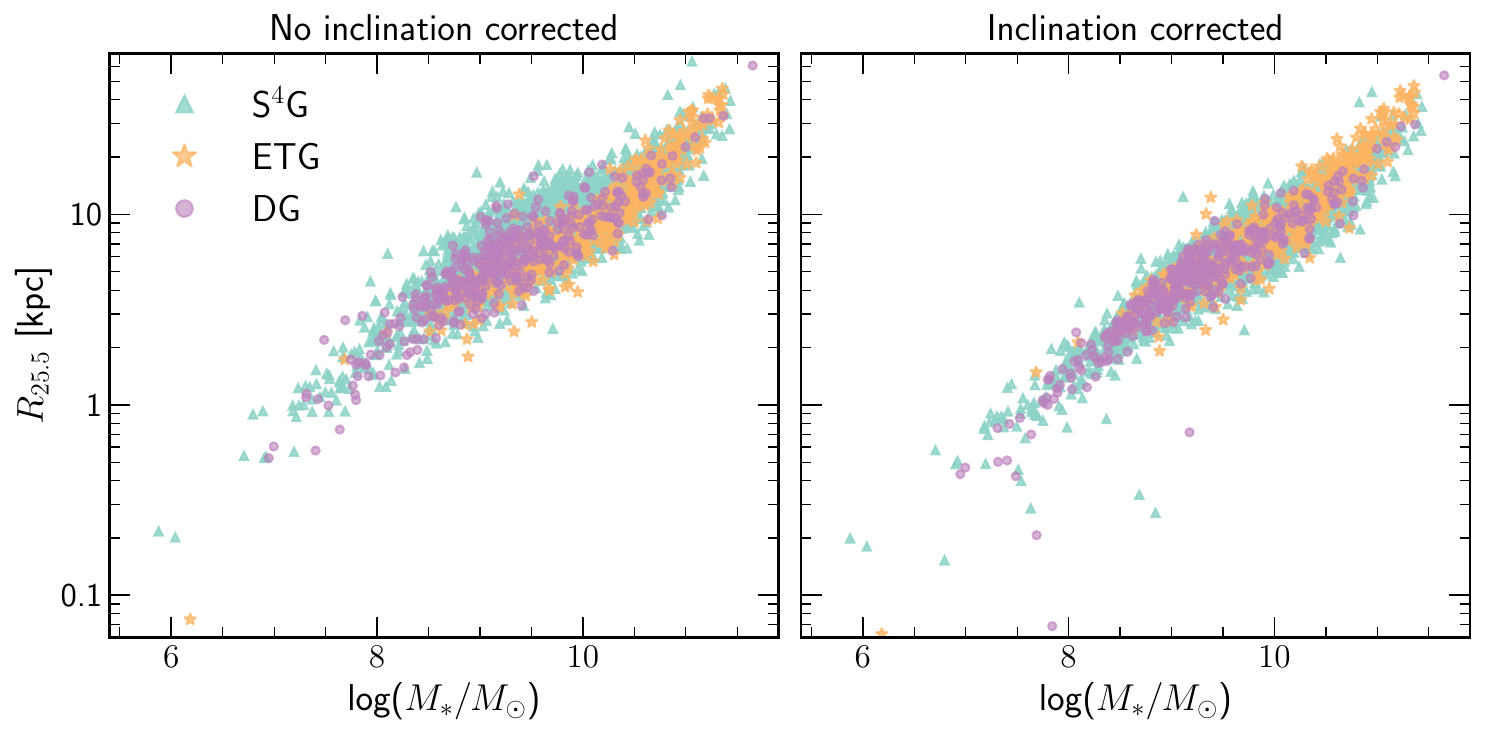}
    \caption{Stellar mass--size relation, showing the correlation of size, traced by the isophotal radius ($R_{25.5}$) with the stellar mass. The DG, ETG, and original S$^4$G samples are represented  with \ltgcolor~\ltgmk, \etgcolor~\etgmk, and \sgcolor~\sgmk respectively. The isophotal radius axis is on a logarithmic scale. The left panel shows the size of galaxies with no correction, and the left panel shows the size of galaxies with inclination correction.}
    \label{fig:Mass_Size}
\end{figure*}

\begin{figure*}[ht!]
    \centering 
    \includegraphics[width=\textwidth]{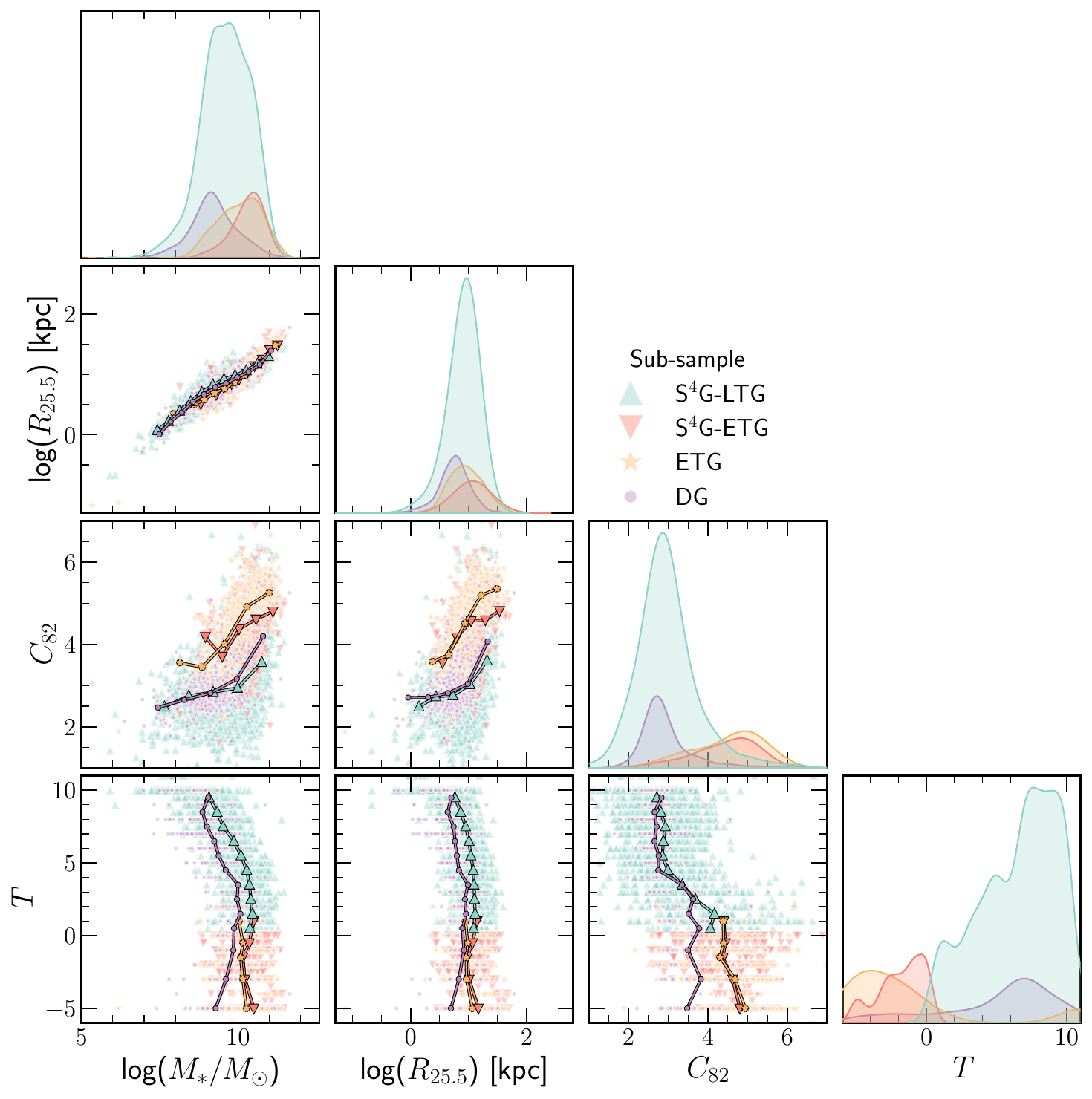}
    \caption{Corner plot of the parameter space described by some of the photometric parameters derived for the DG sample, together with the original sample and ETG extension. From top-left to bottom-right, the diagonal of the figure shows the absolute distribution of the stellar mass, $\log(M_*/M_\odot)$, the galaxy size measured by the isophotal radii at $25.5\,{\rm mag\,arcsec}^{-2}$, $\log(R_{25.5}$), the concentration index, $C_{82}$, and the morphological types, $T$, for the different samples, the original S$^4$G-LTGs (\sgcolor~\sgmk) and S$^4$G-ETGs (\sgetgcolor~\sgetgmk), the ETG extension (\etgcolor~\etgmk), and the DG extension (\ltgcolor~\ltgmk). These distributions are heavily smoothed. Each column and row represents one of the parameters. The first column (left) shows, from top to bottom, galaxy size, concentration index and morphological type versus stellar mass. The second column shows the concentration index and morphological type versus galaxy size. The last column shows morphological types versus concentration index. Curves represent the average values within mass bins for the middle two rows and within morphological types bins for the last row.}
    \label{fig:PairPlot}
\end{figure*}

\section{Discussion}
\label{sec:Scaling}
We present a collection of scaling relations using the derived photometric parameters for the DG extension, in comparison with the original sample \citepalias{Munoz-Mateos15} and the ETG extension \citepalias{Watkins22}. We begin with the mass--size relation, then we explore the parameter space defined by the photometric parameters and the regions where the different morphological types of galaxies reside. We finish by exploring the difference in H\,{\sc i} content for the different samples. 

\subsection{Mass--size relation}
In the left side panel of Fig.~\ref{fig:Mass_Size} we plot the isophotal radius at 25.5\,mag$\,$arcsec$^{-2}$ ($R_{25.5}$) against the stellar mass of the galaxy for the DG extension (\ltgcolor~\ltgmk), the ETG extension (\etgcolor~\etgmk) and the original sample (\sgcolor~\sgmk). \citetalias{Munoz-Mateos15} and \citetalias{Watkins22} showed this relation (in their Figs.~14 and 11, respectively), demonstrating the expected monotonic trend, where galaxies with a higher stellar mass are also larger. Our photometric measurements, properly converted to 3.6$\,\mu$m, reproduce this trend, with excellent agreement between our new galaxies and the original S$^4$G and the ETG samples. The DG sample populates, in general, the intermediate-low mass region ($7\lesssim \log ( M_*/M_\odot)\lesssim11$) with a few massive galaxies with stellar mass above $10^{11}\,M_\odot$. 

The right panel of Fig.~\ref{fig:Mass_Size} shows the mass–-size relation using inclination-corrected isophotal sizes (see Sect.~\ref{sec:Size}). This correction reduces the root mean square deviation of the sizes from the average trends in mass bins by 0.014\,dex, leading to a tighter relation with lower dispersion. This behaviour was also discussed in \citetalias{Watkins22} (their Fig~11), but the corrected values were not published. However, for a few edge-on galaxies, the correction introduces discrepancies, increasing their deviation from the expected trend.\\

Intriguingly, the DG sample contains six large and massive galaxies: NGC~1316, NGC~1404 NGC~4125, NGC~4552,  NGC~7172, and NGC~7410. They can be identified in the top-right corner of Fig.~\ref{fig:Mass_Size}, with size $R_{25.5}>15\,$kpc and stellar mass $\log (M_*/M_\odot)>10.9$. The largest galaxy is NGC~7410 with $R_{25.5}=51\,$kpc and $\log (M_*/M_\odot)=11.36$ while the most massive is NGC~1316 with $R_{25.5}=43\,$kpc and $\log (M_*/M_\odot)=11.65$. The galaxies NGC~1316 and NGC~1404 are part of the Fornax Cluster \citep[see e.g.][and follow-up papers]{Fornax-Ferguson88}. These galaxies follow the same trend as their counterparts in the S$^4$G and ETG samples. 

We also find three dwarf galaxies with mass below  $\log (M_*/M_\odot)< 7$ and smaller than $R_{25.5}<0.6\,$kpc. These galaxies are  PGC~3097691, UGC~04879, and UGC~08308. 

These galaxies underscore the significance of incorporating the DG extension to improve the completeness throughout the mass range of the sample.


\begin{figure*}[ht]
    \centering 
    \includegraphics[width=\textwidth]{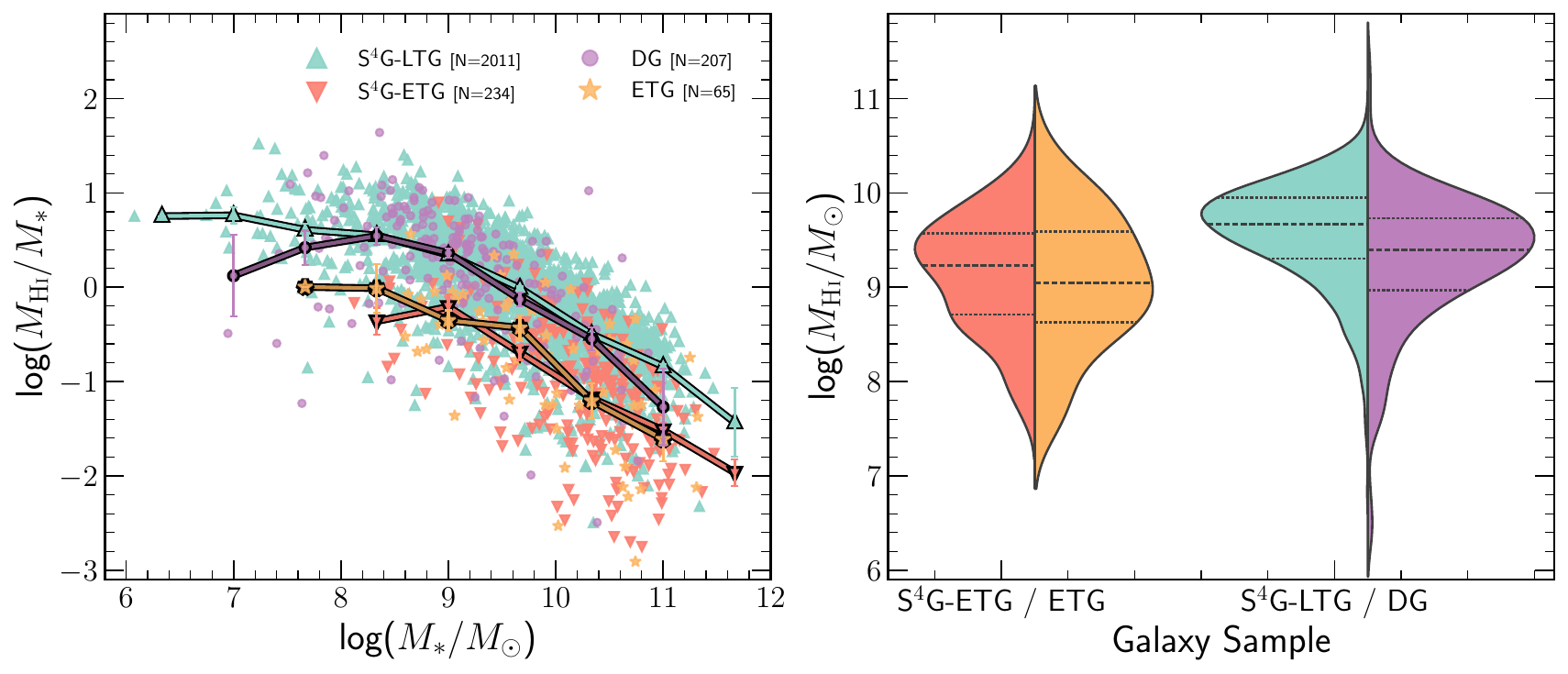}
    \caption{H\,{\sc i} mass content distribution. The left panel shows the logarithmic gas fraction, $\log(M_{\rm H\,\textsc{i}}/M_*)$, versus the stellar mass $\log(M_*/M_\odot)$. The right column shows a violin plot of the H\,{\sc i} mass relative distributions for all the sub-samples. The \textit{x}-axis shows the relative distribution of all sub-samples. The lines show the median values (dashed) and the 25\% and 75\% percentiles (dotted). The S$^4$G-LTG, S$^4$G-ETG, ETG, and DG sub-samples are shown with \sgcolor~\sgmk, \sgetgcolor~\sgetgmk, \etgcolor~\etgmk, and \ltgcolor~\ltgmk, respectively. The numbers shown in the legend of the left panel indicate the number of galaxies with H\,{\sc i} measurements in each sub-sample. }
    \label{fig:HI}
\end{figure*}

\subsection{General trends}
In Fig.~\ref{fig:PairPlot}, we present a corner~plot of the different photometric parameters derived in this work to explore the different relationships between the parameters in a compact way. We defined two subsamples from the S$^4$G, the late-type (S$^4$G-LTG with $T>0$) and the early-type (S$^4$G-ETG with $T<0$) to compare with the DG and ETG extensions. We represent the different parameter spaces defined by the stellar mass, $\log (M_* /M_\odot)$, the size of the galaxy measured as the isophotal radius at 25.5 mag\,arcsec$^{-2}$, $\log (R_{25.5}$), the concentration index, $C_{82}$, and the morphological type, $T$, for the different samples: the original S$^4$G-LTGs (\sgcolor~\sgmk) and S$^4$G-ETGs (\sgetgcolor~\sgetgmk), the ETG extension (\etgcolor~\etgmk), and the DG extension (\ltgcolor~\etgmk). The curves represent the average values per mass bin in the middle two rows and per morphological-type bins in the bottom row.

Along the diagonal of the figure, the panels (from top-left to bottom-right) show the distribution of stellar mass, isophotal radius, concentration index, and morphological type, respectively, for the four different samples. The sample size difference immediately stands out. While the ETG extension (465) is larger than the number of ETG galaxies in the original S$^4$G (282), the DG sample (422) is much smaller than the LTG part of the original S$^4$G (2070), yet it is essential for the completeness of the S$^4$G. 

The stellar masses show different trends in the histogram, with distribution means and standard deviations of $9.23\pm0.78$, $9.77\pm0.83$, $10.11\pm0.82$, and $10.46\pm0.64$  $\log (M_* /M_\odot)$ for the DG, S$^4$G-LTG, ETG, and S$^4$G-ETG, respectively. While the S$^4$G-ETG and ETG have broader distributions, the DG and S$^4$G-LTG distributions are narrower and more skewed. The DG and the ETG sub-samples are 0.54\,dex and 0.35\,dex, respectively, less massive on average than the LTG and ETG sub-samples of the original S$^4$G.

The stellar mass panels (first column) indicate the stellar mass--size relation (second row, same as Fig.~\ref{fig:Mass_Size}), the concentration indices (third row), and the morphological types (fourth row) with respect to the stellar mass. As discussed previously, the parameters derived in this analysis for the DG sample, and properly corrected to the 3.6$\,\mu$m band, follow similar trends as the S$^4$G and ETG. Lower-mass galaxies show smaller radii, lower concentration indices, and higher values of morphological types, while massive galaxies have larger sizes and broader distributions in concentration index and morphological types. These trends are easily seen with the average curves. 

The sizes of galaxies show narrow and aligned distributions for all the sub-samples with some difference in the logarithmic means and standard deviations,  $\log(R_{25.5}/\mathrm{kpc})=0.74\pm0.35$, $0.97\pm0.27$, $0.98\pm0.31$, and $1.11\pm0.27$ for the DG, S$^4$G-LTG, ETG, and S$^4$G-ETG sub-samples, respectively. In units of kiloparsecs, this is $7\pm5$, $11\pm6$, $11\pm9$, and $14\pm8.\,$kpc, respectively. The galaxies in the DG and ETG extension are, on average, 35\% smaller than the LTG of the original S$^4$G, while the galaxies in the ETG extension are 20\% larger. 

The size-related panels (second column) show the concentration indices (third row), and the morphological types (fourth row) with respect to the isophotal radii. Again, as galaxy size is well correlated with stellar mass, we find expected trends, as previously reported by \citetalias{Munoz-Mateos15} and \citetalias{Watkins22}. Smaller galaxies, i.e. less massive, have lower values of $C_{82}$ and later types. However, these relations are less strongly correlated than the ones with respect to stellar mass.

The concentration indices $C_{82}$ (third row) show different distributions (in the third column) for the DG and ETG sub-samples but very similar ones for the extensions and the original S$^4$G. The mean values for each sub-sample are $2.93\pm0.74$, $3.12\pm0.74$, $4.31\pm0.99$, and $4.19\pm0.97$ for the DG, S$^4$G-LTG, ETG, and S$^4$G-ETG, respectively. These differences are less significant than 10\%. However, in general, these figures and average lines show two clear populations that can be identified as LTGs and ETGs, with lower concentration and higher concentration, respectively. As previously reported in many works (see, e.g. \citetalias{Munoz-Mateos15}, \citetalias{Watkins22}, and references therein), the light concentration of a galaxy correlates with its morphological type, and it is one of the parameters typically used for galaxy classification for higher-redshift and unresolved objects. However, among the disc galaxies, there is a notable scatter due to S0 galaxies having both high and intermediate values of concentration indices. 

The morphological type distributions (fourth row) show the variety of galaxies in each sub-sample. All samples combined show a homogeneous and representative of the population in the local Universe. The mean value and standard deviations of $T$-types for each sub-sample are $4.44\pm4.3$, $6.54\pm2.77$, $-1.25\pm5.17$, and $-1.8\pm1.58$ for the DG, S$^4$G-LTG, ETG, and S$^4$G-ETG, respectively. These average values are expected and a consequence of the definition of each sub-sample. We find that for similar morphological types, the galaxies in the DG and ETG extensions are less massive, smaller and less concentrated than their counterparts in the original S$^4$G, as shown in the average values. The plot of the morphological types with respect to the concentration index (third column) also shows the expected clear transition between less-concentrated LTGs and highly concentrated ETGs. 

Our recipes for the conversion between optical \textit{i}-band and 3.6\,$\mu$m band yield results consistent with the original S$^4$G sub-sample and the ETG extension observed with \textit{Spitzer}, as shown in all the panels of Fig~\ref{fig:PairPlot}. By adding this new extension, we increase the sample by 422 new galaxies, which represents 18\% of the original S$^4$G, and 36\% of the sub-sample with mass below $\log(M_*/M_\odot)<9$. Together with the ETG extension, the survey increased the number of galaxies by 38\%, significantly improving its completeness.  

\subsection{Gas content}

Since the DG sub-sample originated from the lack of radio-derived velocities, we may expect that galaxies in this sub-sample exhibit a low H\,{\sc i} content. This particularity raises another interesting reason to study this sub-sample and include it in the CS$^4$G. 

We show the distribution of the H\,{\sc i} mass fraction, $\log(M_{\rm H\,\textsc{i}}/M_*)$, with respect to the stellar mass, $\log(M_*/M_\odot)$, for all the sub-samples in the left panel of Fig.~\ref{fig:HI}. We use the magnitude of the 21\,cm line from HyperLeda to estimate the mass of  H\,{\sc i}. There are 188 (44\%), 2011 (97\%), 65 (14\%), and 234 (83\%) available measurements for the DG, S$^4$G-LTG, ETG, and S$^4$G-ETG sub-samples, respectively. We follow a similar procedure as \cite{1997ApJ...490..173Z, 2023MNRAS.521.5177N} to obtain H\,{\sc i} masses. We measure H {\sc i} masses as follows, 
\begin{equation}
M_{\mathrm{H\,\textsc{i}}}=2.36 \times 10^5\,d^2\,F_{21} \times \frac{1}{1+z} \, ,
\end{equation}
where $d$ is the distance to the galaxy in Mpc, $F_{21}$ is the  21\,cm line integrated flux in Janskys reported by HyperLeda, and $z$ is the redshift of the galaxy. The DG, S$^4$G-LTG, ETG, and S$^4$G-ETG sub-samples are shown in \ltgcolor~\ltgmk, \sgcolor~\sgmk, \etgcolor~\etgmk, and \sgetgcolor~\sgetgmk, respectively. The lines represent the average values within the same bins of stellar mass. The right column shows a violin plot with the H\,{\sc i} mass $\log(M_{\rm H\,\textsc{i}}/M_\odot)$ distribution for all the sub-samples. 

Despite the lack of observational H\,{\sc i} data in much of the DG sub-sample, the average values of the DG galaxies are below the averages of their counterparts in the original S$^4$G-LTG. The disc galaxies in our new DG sub-sample, with the same mass as their LTG counterparts in the original S$^4$G, have a lower H\,{\sc i} mass fraction. The ETG and S$^4$G-ETG sub-samples show no clear difference in their average values. However, only 14\% of the galaxies in the ETG sample have H\,{\sc i} measurements. If these are the ETG galaxies richest in gas, they might represent the upper limit of H\,{\sc i} mass, while the remaining ones might have a lower content of gas. These ETG galaxies most probably are borderline galaxies between ellipticals and spirals. In the right panel, we see a similar trend where the distribution of the DG sub-sample is centred at lower values of H\,{\sc i} mass than the S$^4$G-LTG, with a longer tail towards the low-mass end while the ETGs and S$^4$G-ETG show similar distributions. 



Both the ETG and DG sub-samples have lower stellar masses and, as discussed in the previous subsection,  lower fractions of H\,{\sc i} content than the S$^4$G counterparts with a larger difference for the DG sub-sample. The exclusion of many of these galaxies from the original S$^4$G may be attributed to their lower mass, lower luminosity, and reduced H\,{\sc i} fraction.  Their lack of radio-derived velocities stems from either insufficiently deep data or a lack of H\,{\sc i} observations at the time the original S$^4$G project was defined.

\section{Conclusions} 
\label{sec:conclusions}
We present the Complete \textit{Spitzer} Survey of Stellar Structure in Galaxies (CS$^4$G), a survey of 3239 galaxies with consistent homogenised photometric parameters. We join the original sample with both the early-type \citepalias[ETG][]{Watkins22} and the disc galaxy (DG, this paper) extensions to deliver one catalogue to the community. Additionally, we homogenise all measurements in the catalogue and include new measurements of effective radius and mean surface brightness levels within the effective radius and within 1\,kpc for the whole CS$^4$G. 

We incorporate  401 disc galaxies and 21 elliptical galaxies into the sample. We release archival and new \textit{i}-band imaging for 367 galaxies of the DG extension and 55 $3.6\,\mu$m and $4.5\,\mu$m band images from the \textit{Spitzer} Heritage Archive. We use 102, 169, 77, and 1 images from the DES, LS, SDSS and \textit{HST}, respectively. We observed 18 galaxies in the \textit{i}-band using the LT and the NTT. We analyse all the images using the original S$^4$G methods \citep[\citetalias{Munoz-Mateos15},][\citetalias{Watkins22}]{Salo15} and derive radial surface brightness profiles, curves of growth, asymptotic magnitudes, stellar masses, effective radii, isophotal radii, and concentration parameters. We derive a recipe to transform optical \textit{i}-band to $3.6\,\mu$m magnitudes using images from the original S$^4$G and optical \textit{i}-band images. Using this recipe we convert \textit{i}-band magnitudes to $3.6\,\mu$m to obtain absolutes parameters. 

The CS$^4$G parameters allow us to study different scaling relations and find the following results. 
\begin{itemize}

    \item[-] We improve the completeness of the survey by adding 422 galaxies to the original sample. This represents 15\% of the total previous sample (S$^4$G+ETG). However, for low-mass galaxies ($M_*<10^{9}\,M_\odot$), this sample represents an increment of 36\%.

    \item[-] We measure all the parameters described for the three samples, S$^4$G, ETG and DG, using the same methods, creating a consistent and homogenised CS$^4$G sample. 

    \item[-] Our recipe for the conversion between optical \textit{i}-band to infrared $3.6\,\mu$m band yields measurements consistent with those from the original survey.

    \item[-] We improve the mask images of the DG by increasing the number of pixels masked by a factor of five in comparison with the original S$^4$G masks. We mask regions 2 mag\,arcsec$^{-2}$ fainter. This does not affect the galaxy integrated magnitudes but results in a 52\% lower error on the background characterisation.

    \item[-] The DG extension consists of galaxies with masses $7\lesssim \log ( M_*/M_\odot)\lesssim11$. It contains six massive galaxies with $\log(M_*)>11$ and three galaxies in a tail of lower-mass dwarf galaxies $\log(M_*)<7$. 

    \item[-] The DG galaxies are, on average, 0.23\,dex less massive and 34\% smaller in size than the LTGs of the original S$^4$G. They have similar concentration indices (within 5\%) and later morphological types. 


    \item [-] The DG sample galaxies show a lower  H\,{\sc i} gas fraction than the LTGs in the original S$^4$G. However, we lack H\,{\sc i} measurements for 86\% and 56\% of the ETG and DG samples. Further radio measurements are needed to confirm this finding. 

    \item[-] The CS$^4$G encompasses at least 99.94\% of the complete sample of nearby galaxies meeting the selection criteria of the S$^4$G in the local Universe.

\end{itemize}

In summary, our measurements and our study of the scaling relations show good agreement with previous studies, yet also highlight specific details that are worthy of further investigations. All images, profiles, and derived parameters are made available via the NASA/IPAC Infrared Science Archive, and the CDS. The CS$^4$G will serve as a local benchmark for comparing higher redshift studies from upcoming surveys, such as \textit{Euclid}, Rubin, Roman, and others.

\section{Data Availability}
\label{sec:DR}
We make the CS$^4$G catalogue available with the derived photometric parameters through an electronic table accessible via the CDS and IRSA, together with the images and profiles of the galaxies in the disc extension. This catalogue contains all the information presented in previous samples homogenised with the methods described together with the additional bands studied in this work, the \textit{i}-band (columns with suffix \texttt{3}, e.g., \texttt{mag3}) and the converted 3.6$\,\mu$m (inserted in columns with suffix \texttt{1}, e.g., \texttt{mag1}) magnitudes. We include some extra parameters ($R_{\rm e}$, $\mean{\mu}_{\rm e}$, $\mean{\mu}_{1\,{\rm kpc}}$), and inclination-corrected isophotal radii ($R_{\rm 25.5,corr}$ and $R_{\rm 26.5,corr}$) derived for the original S$^4$G and ETG that were not previously published. They are explained in Sect.~\ref{sec:P3}. We include all the parameters queried from the HyperLeda database. We include the CVRHS morphological classification for the DG extension and for the S$^4$G \citep{Buta15} and ETG samples \citepalias{Watkins22}. 

While the data is being uploaded to IRSA and CDS the provisional link is:
\url{https://cloud.iac.es/index.php/s/3s2qPX6feBJgREg}.

\begin{acknowledgements}
This work is based in part on observations made with the Spitzer Space Telescope, which was operated by the Jet Propulsion Laboratory, California Institute of Technology under a contract with NASA. Based on observations collected at the European Organisation for Astronomical Research in the Southern Hemisphere under ESO programme(s) PPP.C-NNNN(R). Based on observations made with the Liverpool Telescope operated on the island of La Palma by Liverpool John Moores University in the Spanish Observatorio del Roque de los Muchachos of the Instituto de Astrofisica de Canarias with financial support from the UK Science and Technology Facilities Council.  
We acknowledge support from the Agencia Estatal de Investigaci\'on del Ministerio de Ciencia, Innovaci\'on y Universidades (MCIU/AEI) under the grants "The structure and evolution of galaxies and their outer regions" and the European Regional Development Fund (ERDF) with references PID2019-105602GBI00/10.13039/501100011033 and PID2022-136505NB-I00/10.13039/501100011033. 
Co-funded by the European Union (MSCA Doctoral Network EDUCADO, GA 101119830 and Widening Participation, ExGal-Twin, GA 101158446). 
SC acknowledges funding from the State Research Agency (AEI) of the Spanish Ministry of Science, Innovation, and Universities under the grant “The relic galaxy NGC 1277 as a key to understanding massive galaxies at cosmic noon” with reference PID2023-149139NB-I00. AEW acknowledges support from the STFC [grant number ST/X001318/1].
This work was authored by an employee of Caltech/IPAC under contract No. 80GSFC21R0032 with the National Aeronautics and Space Administration. 
LCH was supported by the National Science Foundation of China (11991052, 12233001), the National Key R\&D Program of China (2022YFF0503401), and the China Manned Space Project (CMS-CSST-2021-A04, CMS-CSST-2021-A06). TK acknowledges support from the Basic Science Research Program through the National Research Foundation of Korea (NRF) funded by the Ministry of Education (No. RS-2023-00240212) and the NRF grant funded by the Korean government (MSIT) (No. 2022R1A4A3031306). EA and AB gratefully acknowledge financial support from the CNES (Centre National d’Études Spatiales, France). DAG was supported by STFC grants ST/T000244/1 and ST/X001075/1. JK acknowledges support from NSF through grant AST-2006600. \\
 We acknowledge the usage of the HyperLeda database (http://leda.univ-lyon1.fr). This research has made use of the NASA/IPAC Extragalactic Database (NED), which is operated by the Jet Propulsion Laboratory, California Institute of Technology, under contract with the National Aeronautics and Space Administration. This research has made use of the SIMBAD database, operated at CDS, Strasbourg, France.
Funding for the Sloan Digital Sky Survey V has been provided by the Alfred P. Sloan Foundation, the Heising-Simons Foundation, the National Science Foundation, and the Participating Institutions. SDSS acknowledges support and resources from the Center for High-Performance Computing at the University of Utah. SDSS telescopes are located at Apache Point Observatory, funded by the Astrophysical Research Consortium and operated by New Mexico State University, and at Las Campanas Observatory, operated by the Carnegie Institution for Science. The SDSS web site is \url{www.sdss.org}. SDSS is managed by the Astrophysical Research Consortium for the Participating Institutions of the SDSS Collaboration, including Caltech, The Carnegie Institution for Science, Chilean National Time Allocation Committee (CNTAC) ratified researchers, The Flatiron Institute, the Gotham Participation Group, Harvard University, Heidelberg University, The Johns Hopkins University, L’Ecole polytechnique f\'{e}d\'{e}rale de Lausanne (EPFL), Leibniz-Institut für Astrophysik Potsdam (AIP), Max-Planck-Institut für Astronomie (MPIA Heidelberg), Max-Planck-Institut für Extraterrestrische Physik (MPE), Nanjing University, National Astronomical Observatories of China (NAOC), New Mexico State University, The Ohio State University, Pennsylvania State University, Smithsonian Astrophysical Observatory, Space Telescope Science Institute (STScI), the Stellar Astrophysics Participation Group, Universidad Nacional Aut\'{o}noma de M\'{e}xico, University of Arizona, University of Colorado Boulder, University of Illinois at Urbana-Champaign, University of Toronto, University of Utah, University of Virginia, Yale University, and Yunnan University. The DESI Legacy Imaging Surveys consist of three individual and complementary projects: the Dark Energy Camera Legacy Survey (DECaLS), the Beijing-Arizona Sky Survey (BASS), and the Mayall z-band Legacy Survey (MzLS). DECaLS, BASS and MzLS together include data obtained, respectively, at the Blanco telescope, Cerro Tololo Inter-American Observatory, NSF’s NOIRLab; the Bok telescope, Steward Observatory, University of Arizona; and the Mayall telescope, Kitt Peak National Observatory, NOIRLab. NOIRLab is operated by the Association of Universities for Research in Astronomy (AURA) under a cooperative agreement with the National Science Foundation. Pipeline processing and analyses of the data were supported by NOIRLab and the Lawrence Berkeley National Laboratory (LBNL). Legacy Surveys also uses data products from the Near-Earth Object Wide-field Infrared Survey Explorer (NEOWISE), a project of the Jet Propulsion Laboratory/California Institute of Technology, funded by the National Aeronautics and Space Administration. Legacy Surveys was supported by: the Director, Office of Science, Office of High Energy Physics of the U.S. Department of Energy; the National Energy Research Scientific Computing Center, a DOE Office of Science User Facility; the U.S. National Science Foundation, Division of Astronomical Sciences; the National Astronomical Observatories of China, the Chinese Academy of Sciences and the Chinese National Natural Science Foundation. LBNL is managed by the Regents of the University of California under contract to the U.S. Department of Energy. The complete acknowledgements can be found at \url{https://www.legacysurvey.org/acknowledgment/}.

\textit{Sotware: } This work made use of \texttt{Astropy} (\url{http://www.astropy.org}) a community-developed core Python package and an ecosystem of tools and resources for astronomy \citep{astropy:2013, astropy:2018, astropy:2022}; \texttt{Photutils}, an Astropy package for detection and photometry of astronomical sources \citep{photutils-v1.5.0}; \texttt{Matplotlib} \citep{matplotlib}; \texttt{NumPy} \citep{numpy}; \texttt{SciPy} \citep{scipy}; \texttt{Pandas} \citep{pandas}; \texttt{TOPCAT} \citep{topcat}; \texttt{SExtractor} \citep{SExtractor}; \texttt{Swarp} \citep{Swarp}; and \texttt{SAO Image DS9} \citep{ds9}. 
\end{acknowledgements}
\bibliographystyle{aa}
\bibliography{bibliography}

\begin{appendix}
\section{Origin of the bias in the S$^4$G sample}
\label{app:sample}

\begin{figure}[b!]
    \centering
    \includegraphics[width=0.49\textwidth]{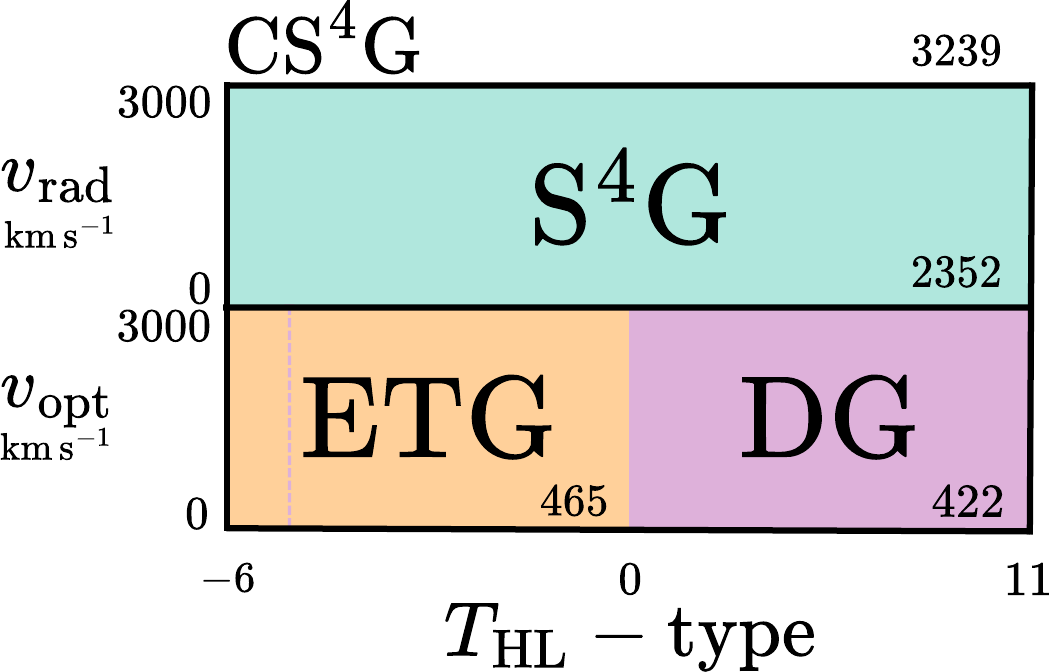}
    \caption{Illustration of the CS$^4$G sample. The CS$^4$G contains the S$^4$G, ETG and DG sample. It mimics the criteria used to define the S$^4$G, but including also optical velocities to define the ETG and DG samples. }
    \label{fig:cs4g_graphic}
\end{figure}

The S$^4$G aimed to construct a volume-, magnitude-, and size-limited survey of galaxies in the local Universe and study the spatial distribution of light through mid-infrared imaging. Originally, when the sample of the S$^4$G was selected \citep{Seth10}, the query of the galaxies in the HyperLeda database returned 2331 galaxies. For this selection, radio H\,{\sc i} velocities as a query for distance were used to select galaxies closer than $40\,$Mpc. However, the use of radio-derived H\,{\sc i} velocities biased the original sample towards gas-rich late-type galaxies and missed a high fraction (over 50\%) of early-type galaxies, which typically have lower gas content.

\cite{Seth13-Spitzer} used HyperLeda values based on optical spectroscopy to determine the distance to galaxies to search for early-type galaxies ($T_\mathrm{HL}\leq0$) without H\,{\sc i} velocities. This introduced 465 new early-type galaxies \citepalias{Watkins22}, lacking radio H\,{\sc i} measurements in the HyperLeda database, which made them miss the criteria of the original S$^4$G but meeting the volume-limited criteria of $40\,$Mpc using optical-derived distances. Nevertheless, the specific criterion to search for early-type galaxies ($T_\mathrm{HL}\leq0$) introduced a new bias against late-type galaxies. 

Later on, when a study of the Fornax cluster with S$^4$G authors involved was being carried out \citep{Venhola17-Fornax}, they found out that many disc galaxies were not in the S$^4$G despite meeting all requisites. This finding made the authors revise the original sample and select disc galaxies using optical velocities, giving rise to the DG extension. As explained in Sect.~\ref{sec:Sample}, the DG extension mimics the criterion of the ETG extension but searches for late-type galaxies ($T_\mathrm{HL}>0$).  Intriguingly, several galaxies with $T_\mathrm{HL}\leq0$ were not included in the early-type extension but are now included in the disc extension. We are not sure why these galaxies were missed in the original or ETG sample, but one possible explanation is that new measurements could have been added in HyperLeda for these galaxies between the time each of the samples was done (2013 and 2018\footnote{Between 2018 and 2024, there have been no significant changes in the parameters of CS$^4$G galaxies in HyperLeda}). Another explanation, at least for galaxies with large apparent size (such as NGC~1316), is that the original S$^4$G excluded them on purpose not to expend many pointings of \textit{Spitzer} on these galaxies. Now, with this publication, we fulfil the original aim of constructing a volume-, magnitude-, and size-limited survey by adding 422 disc galaxies to the sample.

We present the CS$^4$G, a combination of the S$^4$G, ETG and DG subsamples, with a final total size of 3239 galaxies in the local Universe. Fig.~\ref{fig:cs4g_graphic} shows an illustration of the CS$^4$G and the differences between the ETG and DG samples. The \textit{x}-axis represents the morphological $T$-types, while the \textit{y}-axis represents the recessional velocity of the galaxies. The ETG and DG samples differ from the S$^4$G by the wavelength used to measure the recessional velocity, optical and radio H\,{\sc i} velocities, respectively. They also differ by the morphological types, while ETG includes $T_\mathrm{HL}\leq0$, the DG mostly includes $T_\mathrm{HL}>0$ with 33 $T_\mathrm{HL}\leq0$ galaxies that missed the ETG (represented with the purple dashed line).  



\section{Galaxies not forming part of the DG sample.}
\label{app:exclusion}
The exact query used in HyperLeda, mimicking the selection criteria of the CS$^4$G is the following:
\begin{lstlisting}[
           language=SQL,
           showspaces=false,
           basicstyle=\ttfamily,
           numberstyle=\tiny,
           commentstyle=\color{gray}
        ]
SELECT pgc,objname,al2000,de2000,logd25,
b2,t,bt,btc,vmaxg,vmaxs,v,vrot,modz,mod0,
modbest,mabs,vrad,vopt,objtype

WHERE (objtype='G' OR objtype='?')  
AND btc<15.5 AND logd25 >1. 
AND logd25<2.48 AND abs(b2)>30. 
AND (vrad <3000 OR vopt <3000)
\end{lstlisting}
This query executed on October 2024\footnote{This is the date of the last execution} resulted in 430 new galaxies not included in the S$^4$G and ETG. However, of these galaxies 8 were rejected because the galaxy either appeared to be misclassified, it was a stellar stream related to another galaxy, a system of interacting galaxies or it was a background galaxy. The galaxies rejected were: PGC~015573, PGC~166170, PGC~3097827, PGC~014117, PGC~014118, PGC~014121, PGC~068481, and ESO~056-019.

PGC~015573 is misclassified as a galaxy. We deduce it must be a star system embedded in a dust cloud that produces an extended emission which could be interpreted as a galaxy in shallower data. PGC~166170, also known as KK~208, is a stellar stream of M83 with an estimated stellar mass of $\sim 1\times10^{8} M_\odot$ \citep{2014ApJ...789..126B}. The galaxy PGC~3097827, also known as F8D1, is an Ultra Diffuse Galaxy (UDG) associated with M81 \citep{2000A&A...363..117K} with a surface brightness peak of 25.81\,mag\,arcsec$^{-2}$ and an integrated magnitude fainter than 15.5\,mag. The galaxies PGC~014117, PGC~014118, and PGC~014121 appear to be systems of interacting galaxies, with individual galaxies having diameters below the limit. PGC~068481 is a bright knot in the galaxy ESO~602-003, which is included in the S$^4$G. Finally, ESO~056-019 is a region of the LMC. 
Rejecting these 8 entries results in a total sample of 422 galaxies in the DG extension.


\section{Notes on the analysis of UGC~07636 and PGC~3097691.}

The galaxy UGC~07636  is a companion of M49 embedded in the galaxy's outskirts. The galaxy M49 is in the original S$^4$G, and its morphological decomposition was done by \cite{Salo15}. We have used the residual S$^4$G image, after subtracting the model fitted by \cite{Salo15}, to measure the radial profile of UGC~07636 properly. The dwarf galaxy PGC~3097691 is also included in the sample. At a distance of $0.8\,$Mpc the Legacy Surveys imaging can resolve the stars, and despite being analysed, the surface brightness technique is not optimal for studying the structure of the object.

\section{Distance selection}
\label{app:distace}
\begin{figure}[b!] 
    \centering
    \includegraphics[width=0.45\textwidth]{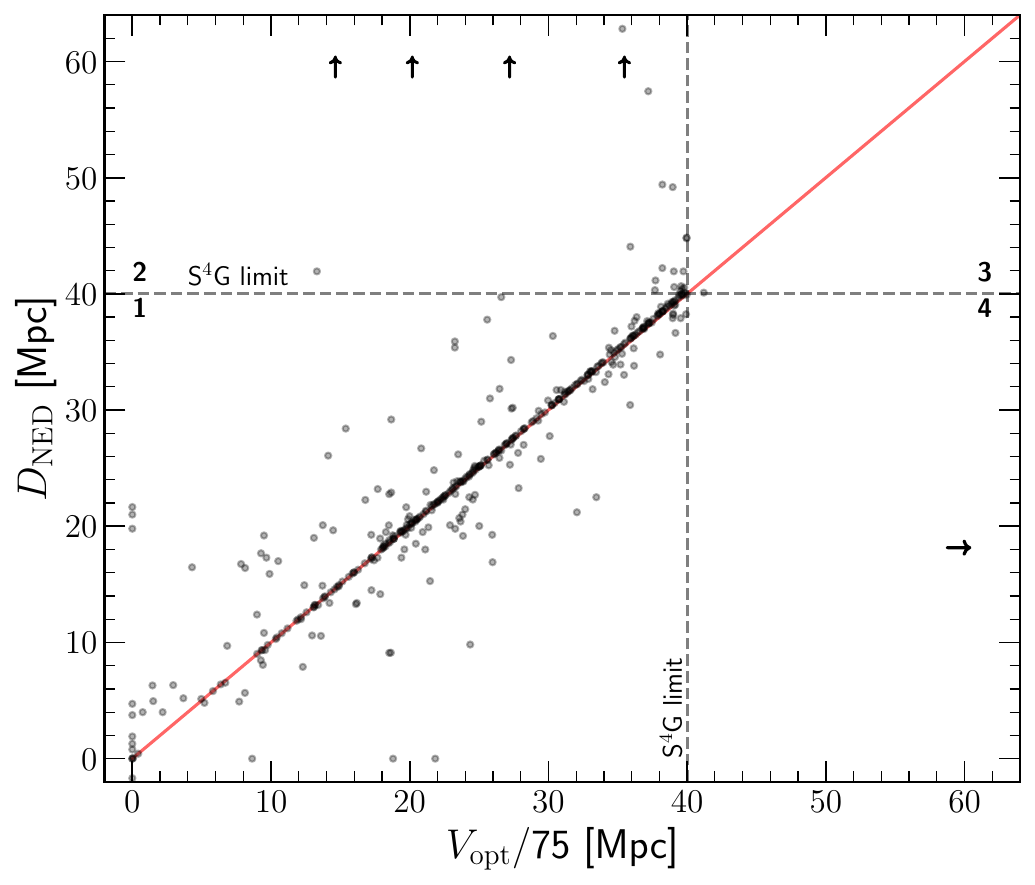}
    \caption{Comparison of distances. The \textit{x}-axis shows the distances obtained using the optical velocities from the HyperLeda database. The \textit{y}-axis shows the distances obtained from the NED database, a combination of redshift-independent measurements and redshift distances. The red line shows the one-to-one relation. The arrow shows galaxies with distances above the limits of the figure but located at the position of the other axis values. The dashed grey lines show the volume limit criterion of the S$^4$G of $40\,$Mpc. }
    \label{fig:distance}
\end{figure}

When selecting the distance to the galaxies to transform apparent to absolute magnitudes, we aim to be consistent with the original S$^4$G, just as we have been with the sample analysis. In the S$^4$G, while the sample selection was based on HyperLeda velocities, the distances were selected from the NED database. A redshift-independent measurement was used in the original S$^4$G, and if that was unavailable, a distance derived from the redshift was used assuming $H_0=75\,$km\,s$^{-1}$\,Mpc$^{-1}$. We use the same methodology. However, there are a few cases where the redshift-independent distance is incompatible with its redshift and others where the NED distances are incompatible with the HyperLeda optical velocities. Fig.~\ref{fig:distance} shows the distances obtained from the HyperLeda radial velocities in comparison with the NED distances.  

The NED distances are selected as explained above and are a combination of redshift-independent measurements and redshift-derived distances. We also show the lines representing the volume limit criterion of the S$^4$G of $40\,$Mpc. These lines divide the figure into four quadrants. The first quadrant, in the bottom left, shows galaxies with distances from the two different sources (NED and HyperLeda) within the criteria of the S$^4$G (below $40\,$Mpc). The third quadrant, in the upper right, as expected, is empty since it represents galaxies not meeting the criterion in either of the sources. The second and fourth quadrants show galaxies meeting the criterion in one of the sources but not in the other. Since the sample selection criterion of the S$^4$G was only based on the HyperLeda database, we would expect not to see any galaxies in the fourth quadrant. In spite of this, we find two galaxies (ESO~288-049 and ESO~603-006) with optical velocities corresponding to distances larger than 40\,Mpc in the HyperLeda database. These two galaxies do have H\,{\sc i} radio velocity measurements below $3000\,$km\,s$^{-1}$. Moreover, there are 25 galaxies with NED distances over the 40\,Mpc criterion, but with optical velocities meeting the selection criterion (seen in the second quadrant). 
For the particular case of the galaxy PGC~040869, we use the redshift-independent distance from \cite{2024ApJ...966..145C}. NED and HyperLeda both register a redshift estimated distance of 0.43\,Mpc. However, the galaxy appears to be one of the many dwarfs around the Virgo Cluster.  \cite{2024ApJ...966..145C} obtained a distance estimation of 13.86\,Mpc using surface brightness fluctuations, which links the galaxy to the Virgo Cluster as expected. 

In summary, our criterion for assigning distances is to ensure they meet the requirement of being less than $40\,$Mpc, regardless of the source. We prioritize our selections in the following order: first, we use NED redshift-independent distances; second, NED redshift distances; and third, optical velocities from HyperLeda. If a measurement from a higher-priority source does not meet the $d<40\,$Mpc criterion, we move to the next priority source.

\section{CVRHS classifications}
The morphological classification and their numerical values (described in Sect.~\ref{subsec:Morphology}) can be found in the electronic table (see Sect.~\ref{sec:DR}) in the columns \texttt{CVRHS} and \texttt{CVRHS\_t}, respectively, and is also shown in Table~\ref{tab:CVRHS}. Table~\ref{tab:morph_t-type} shows the transformation from the morphological classification to numerical values. 

\begin{table}[h!]
    \centering
    \caption{Numerical $T$-types transformation.}
    \begin{tabular}{lc}
    \hline \hline 
    \noalign{\smallskip}
    Stage & Numerical index ($T$-type) \\
    \noalign{\smallskip}
    \hline 
    \noalign{\smallskip}
    cE & -6 \\
    E & -5 \\
    E$^{+}$ & -4 \\
    S$0^{-}$ & -3 \\
    S$0$ & -2 \\
    S$0^{+}$ & -1 \\
    S$0 / \mathrm{a}$ & 0 \\
    Sa & 1 \\
    Sab & 2 \\
    Sb & 3 \\
    Sbc & 4 \\
    Sc & 5 \\
    Scd & 6 \\
    Sd & 7 \\
    Sdm & 8 \\
    Sm & 9 \\
    Im & 10 \\
    $\mathrm{dE}, \mathrm{dS} 0, \mathrm{dSph}$ & 11 \\
    \hline
    \end{tabular}   
    \tablebib{\citet{deVaucouleurs77, Buta15}; \citetalias{Watkins22}}
    \label{tab:morph_t-type}
\end{table}

\onecolumn
\begin{longtable}{llcllc}

\caption{CVRHS classification of the disc galaxy sample. The description of all symbols can be found in \cite{Buta15}. The symbol K stands for galaxies showing a knotted outer ring.}
\label{tab:CVRHS}\\
\hline \hline
\noalign{\smallskip}
Name & CVRHS & $T$ & Name & CVRHS & $T$ \\
\noalign{\smallskip}
\hline
\noalign{\smallskip}	
\endhead

\hline
\endfoot

ESO~013-009 & SAB(s)cd &   6                                                 & ESO~299-011 & SB(s)d &   7                                               \\     
ESO~013-021 & SB$_\textrm{a}$(rs)0$^+$ &  -1                                 & ESO~301-011 & (R)K / S0/a & 0                                            \\
ESO~027-021 & (R$^{\prime}$)SA(nr)b &   3                                    & ESO~306-013 & I0 / IAB(s)m & 0                                          \\
ESO~028-007 & SAB(s)cd &   6                                                 & ESO~340-012 & SA(s)cd &   6                                              \\
ESO~030-008 & Sb: sp &   3                                                   & ESO~340-019 & Sd sp &   7                                                \\
ESO~078-022 & SB(s)cd sp &   6                                               & ESO~340-021 & (R$^{\prime}$)SA(r\underline{s})b &   3                    \\
ESO~079-002 & SAB(s)d &   7                                                  & ESO~345-002 & SA(r$^{\prime}$l)d &   7                                   \\
ESO~080-006 & SA\underline{B}(rs)m &   9                                     & ESO~345-050 & SBd sp &   7                                               \\
ESO~084-015 & SA(s)cd &   6                                                  & ESO~346-018 & SA(s)cd &   6                                              \\
ESO~085-065 & S0bc: sp &   4                                                 & ESO~347-002 & Sd sp &   7                                                \\
ESO~108-023 & SAB(r\underline{s})c &   5                                     & ESO~347-020 & SAc: &   5                                                 \\
ESO~114-032 & SAB(s)dm sp &   8                                              & ESO~349-039 & SB(s)d &   7                                               \\
ESO~119-005 & Im sp &  10                                                    & ESO~352-036 & SAB(rs)d: &   7                                            \\
ESO~119-059 & S\underline{A}B(r\underline{s})0$^\textrm{o}$ &  -2            & ESO~356-026 & SB(s:)dm: &   8                                            \\
ESO~149-013 & SB(s)dm &   8                                                  & ESO~359-013 & E(d)2-3 &  -5                                              \\
ESO~151-019 & (R$^{\prime}$)S\underline{A}B(s)d &   7                        & ESO~400-023 & LTG & -                                                    \\
ESO~153-019 & (R$_2^{\prime}$)SB(s)d &   7                                   & ESO~400-026 & (R$^{\prime}$:)SBd   & 7                                            \\
ESO~157-044 & SAB(rs)m &   9                                                 & ESO~400-027 & Sbc:  & 4                                                      \\
ESO~158-003 & SB(s)dm &   8                                                  & ESO~404-028 & SA(rs)ab &   2                                             \\
ESO~158-015 & S\underline{A}B(s)dm &   8                                     & ESO~404-039 & SA(\underline{r}l)0/a & 8                                          \\
ESO~186-062 & SB(r\underline{s})c &   5                                      & ESO~405-014 & Sdm sp &   8                                               \\
ESO~189-023 & Sd sp &   7                                                    & ESO~406-031 & Sbc: sp / E(d)7 &   4                                      \\
ESO~190-011 & SB(s)d &   7                                                   & ESO~407-004 & SAB(s)dm &   8                                             \\
ESO~200-053 & (R$^{\prime}$L)SB(\underline{r}s)0$^+$ &  -1                   & ESO~410-005 & dIm &  10                                                  \\
ESO~200-054 & SA(s)bc &   4                                                  & ESO~411-016 & S\underline{A}B(s)cd: &   6                                \\
ESO~233-044 & SB(rs)cd: &   6                                                & ESO~411-027 & SB(s)dm &   8                                              \\
ESO~233-053 & SB(rs)c: &   5                                                 & ESO~420-005 & SA(rs)c &   5                                              \\
ESO~234-023 & SB(s)\underline{d}m sp &   7.5                                 & ESO~439-025 & Sc: sp & 5                                                  \\
ESO~234-036 & (R$^{\prime}$)S\underline{A}B(s)dm &   8                       & ESO~440-012 & SAB(s)c &   5                                              \\
ESO~235-001 & Sd sp &   7                                                    & ESO~440-018 & SA(s)c &   5                                               \\
ESO~236-006 & (R$_2^{\prime}$)SB(rs)c\underline{d} &   6.5                   & ESO~443-008 & S\underline{A}B(s)c &   5                                  \\
ESO~236-034 & SAB(s)dm &   8                                                 & ESO~443-033 & SB$_\textrm{a}$0$^-$ &  -3                                 \\
ESO~236-035 & S\underline{A}B(s)c\underline{d} &   6.5                       & ESO~444-084 & dIAm &  10                                                 \\
ESO~236-036 & SA(rs)c pec &   5                                              & ESO~445-076 & SA(s)d pec &   7                                           \\
ESO~236-041 & SB(s)dm &   8                                                  & ESO~462-028 & SA\underline{B}(s)m &   9                                  \\
ESO~237-019 & SABm & 9                                                          & ESO~462-032 & Sd sp / E(d)8 &   7                                        \\
ESO~238-005 & SA(s)d &   7                                                   & ESO~466-025 & S0$^-$ sp / E(d)5 &  -3                                    \\
ESO~240-012 & SA(rs)d pec &   7                                              & ESO~466-043 & Sb sp / E7 &   3                                           \\
ESO~241-006 & SAB(r$^{\prime}$l,ns)b &   3                                   & ESO~468-006 & SA(s)m &   9                                               \\
ESO~251-012 & (R$^{\prime}$)Sdm &   8                                        & ESO~468-008 & IB(s)m &  10                                               \\
ESO~252-007 & SB(rs)d &   7                                                  & ESO~477-005 & SAB(s)cd &   6                                             \\
ESO~284-047 & SA(rs)0$^+$: &  -1                                             & ESO~480-017 & (R$^{\prime}$)SB(s)cd &   6                                \\
ESO~284-049 & (R$_1^{\prime}$)SB(s)cd &   6                                  & ESO~482-017 & SA(s)cd &   6                                              \\
ESO~286-063 & (L)SAB(r\underline{s})c &   5                                  & ESO~482-036 & (R$^{\prime}$)S\underline{A}B(s)cd &   6                   \\
ESO~287-055 & SAB(s)cd &   6                                                 & ESO~482-049 & Im sp / E5 &  10                                           \\
ESO~288-005 & S(\underline{r}s)dm  & 8                                       & ESO~503-008 & Scd spw / E(d)7 &   6                                      \\
ESO~288-025 & Sc sp / E(d)8 &   5                                            & ESO~504-030 & (R$^{\prime}$)SB(r\underline{s})cd &   6                   \\
ESO~288-045 & SB(\underline{r}s)d &   7                                      & ESO~505-005 & (L)SB(s)m &   9                                            \\
ESO~288-049 & SB(s)d &   7                                                   & ESO~508-003 & SA(rs)m &   9                                              \\
ESO~289-005 & (R$^{\prime}$L)SAB(s)d &   7                                   & ESO~508-004 & RK sp & -                                                  \\
ESO~289-011 & SAB(s)cd &   6                                                 & ESO~508-066 & SAB(s)dm &   8                                             \\
ESO~289-020 & SAB:(r:)cd sp &   6                                            & ESO~509-064 & SAB(s)m / I0 pec &   9                                     \\
ESO~289-042 & SB(s) & -                                                      & ESO~510-043 & SAB(rs)cd &   6                                            \\
ESO~290-006 & (R$^{\prime}$)SA(s)bc &   4                                    & ESO~532-015 & Sm sp &   9                                                \\
ESO~290-039 & SAB(s)dm &   8                                                 & ESO~533-005 & SAa sp / E(d)0 &   1                                       \\
ESO~291-003 & Sd sp &   7                                                    & ESO~533-047 & Sb: sp & 3                                                 \\
ESO~294-020 & SA(s)cd &   6                                                  & ESO~543-023 & SAB(s)dm sp &   8                                          \\
ESO~297-015 & SB(s)dm: &   8                                                 & ESO~546-033 & SA(s)dm &   8                                              \\
ESO~297-020 & SB(s)d &   7                                                   & ESO~548-022 & Sc: sp & 5                                                 \\
ESO~548-029 & SB(s)dm &   8                                                  & NGC~2856 & (RL)SAB(rs,r)a: pec &   1                                     \\
ESO~548-034 & SA\underline{B}(rs)c &   5                                     & NGC~2880 & SAB0$^-$ &  -3                                                \\
ESO~548-049 & dIm &  10                                                      & NGC~3230 & SA\underline{B}$_\textrm{a}$(s)0$^\textrm{o}$ &  -2           \\
ESO~548-065 & Sd sp / E7-8 &   7                                             & NGC~3245 & SA(l)0$^\textrm{o}$ &  -2                                     \\
ESO~548-070 & SA(s)cd sp / E8 &   6                                          & NGC~3367 & SB(\underline{r}s)bc &   4                                    \\
ESO~548-073 & SA\underline{B}(s)d &   7                                      & NGC~3400 & S\underline{A}B(\underline{r}s)0/a &   0                      \\
ESO~548-079 & E1-2 &  -5                                                     & NGC~3418 & SAB(s)0/a &   0                                               \\
ESO~551-020 & S0d sp / E(d)7 &   7                                           & NGC~3524 & SA(l)0$^-$ &  -3                                              \\
ESO~572-011 & SAB(s)c &   5                                                  & NGC~3543 & SB(s)m sp &   9                                               \\
ESO~575-023 & SA\underline{B}(\underline{r}s,bl)a &   3                      & NGC~3769A & IB:m: &  10                                                  \\
ESO~576-037 & SB(s)d sp &   7                                                & NGC~4027A & SB(rs)dm &   8                                               \\
ESO~580-005 & SBd spw pec &   7                                              & NGC~4108A & SA\underline{B}(s)cd &   6                                   \\
ESO~601-001 & Im sp / E(d)6-7 &  10                                          & NGC~4125 & E(d)0-1 & -5                                                  \\
ESO~603-005 & SB(s)d sp / E(d)7 &   7                                        & NGC~4218 & IAB(s)m &  10                                                 \\
IC~0127 & Sb sp / E7 &   3                                                   & NGC~4300 & SA(r)0$^+$ &  -1                                              \\
IC~0277 & (R$_1^{\prime}$)SA\underline{B}(rs,nr)b &   3                      & NGC~4305 & SAB(s)a &   1                                                 \\
IC~0540 & Sab sp / E(d)7-8 &   2                                             & NGC~4332 & (R$^{\prime}$)SA\underline{B}(s)a &   1                       \\
IC~0559 & SB(s)d: pec &   7                                                  & NGC~4431 & SAB0$^\textrm{o}$ & -2                                                 \\
IC~0782 & SA(r)0$^\textrm{o}$ &  -2                                          & NGC~4433 & (R$^{\prime}$L)SAB(rs)b &   3                                 \\
IC~1980 & SA(rs)c: &   5                                                     & NGC~4440 & SB$_\textrm{a}$(r\underline{s},bl)a &   1                     \\
IC~2009 & SAB(s)m &   9                                                      & NGC~4469 & SB$_\textrm{x}$a sp pec &   1                                 \\
IC~2038 & Sd: sp / E(d)6-7 &   7                                             & NGC~4474 & S0$^\textrm{o}$ sp / E(d)2-3 &  -2                            \\
IC~2049 & SA(s)cd &   6                                                      & NGC~4481 & SA(rs)bc: &   4                                               \\
IC~3077 & E3 pec &  -5                                                       & NGC~4500 & (R$_1^{\prime}$)SAB(\underline{r}s,nl)ab &   2                \\
IC~3118 & (R$^{\prime}$)SAB(rs)d: &   7                                      & NGC~4516 & SB(s)0/a &   0                                                \\
IC~3430 & dSA(r)0$^\textrm{o}$ &  -2                                         & NGC~4552 & E0-1 & -5                                                     \\
IC~3459 & S0$^-$ sp / E0 &  -3                                               & NGC~4626 & SB(s)dm &   8                                                 \\
IC~3471 & SA(rs)cd: &   6                                                    & NGC~4674 & SAab sp / E(b)6 &   2                                         \\
IC~3518 & S0$^-$ sp / E(b)6 &  -3                                            & NGC~4964 & SA(s)c &   5                                                  \\
IC~3530 & E2-3 (isophote twisting)) &  -5                                    & NGC~5037 & (R$^{\prime}$)SA(s:)b &   3                                   \\
IC~3647 & dS0(5) &  11                                                       & NGC~5294 & IBm &  10                                                     \\
IC~3720 & dE6-7 &  11                                                        & NGC~5323 & (R)SA(s)b spw / E0 &   3                                      \\
IC~4249 & S0$^+$? pec &  -1                                                  & NGC~5378 & (R$^{\prime}$)SB(r,bl)ab &   2                                \\
IC~4323 & Sd spw / E8 &   7                                                  & NGC~5475 & S0$^-$ sp / E(d)7 &  -3                                       \\
IC~4336 & SA(rs)b &   3                                                      & NGC~5705 & SB(s)d &   7                                                  \\
IC~4946 & (RL)SB$_\textrm{x}$0/a &   0                                       & NGC~5811 & SB(s)m &   9                                                  \\
IC~5028 & I\underline{A}Bm &  10                                             & NGC~5866 & S0$^-$ sp / E4 &  -3                                          \\
IC~5162 & S\underline{A}B:m pec &   9                                                & NGC~5917 & (R)SB(rs)d &   7                                              \\
IC~5224 & I0 sp pec & 0                                                      & NGC~6186 & (R$^{\prime}$L)SB(rs)a &   1                                  \\
IC~5267A & SB(s)d &   7                                                      & NGC~6283 & SA(rs)d &   7                                                 \\
NGC~0172 & Sc sp/ E(d)7 &   5                                                & NGC~6306 & Sb sp / E(d)7 &   3                                           \\
NGC~0586 & SA(rs)ab &   2                                                    & NGC~6861B & S0$^\textrm{o}$  / E(d)6 &  -2                               \\
NGC~0853 & (R$^{\prime}$L)SAB(s)dm &   8                                     & NGC~6890 & (R$^{\prime}$L)SAB(rs)a &   1                                 \\
NGC~1189 & SB(s)d &   7                                                      & NGC~6990 & SA(\underline{r}s)cd &   6                                    \\
NGC~1253A & S/IAB:(s:)m &  10                                                & NGC~7096 & SA(rs)ab &   2                                                \\
NGC~1266 & (RL)SA(rs)0/a &   0                                               & NGC~7126 & SA(rs)bc &   4                                                \\
NGC~1316 & E$^+$3 &  -4                                                      & NGC~7172 & E4 (twisted to E2-3) &  -5                                    \\
NGC~1317 & (R$^{\prime}$)S\underline{A}B(rl,nr,nb)0/a &   0                  & NGC~7174 & E0 & -5                                                       \\
NGC~1320 & E6: &  -5                                                         & NGC~7199 & (RR\underline{R})SAB(r)a &   1                                \\
NGC~1369 & S\underline{A}B(rs)a &   1                                        & NGC~7232 & Sb sp / E(d)7 &   3                                           \\
NGC~1377 & SA0$^-$ &  -3                                                     & NGC~7232A & (R$^{\prime}$)SA(rs)b: &   3                                 \\
NGC~1404 & E1 &  -5                                                          & NGC~7232B & SB(s)cd &   6                                                \\
NGC~1522 & (L)Im &  10                                                       & NGC~7233 & (R$^{\prime}$)SAB(\underline{r}s)a pec &   1                  \\
NGC~1536 & SB(s)dm &   8                                                     & NGC~7259 & SA(s)bc &   4                                                 \\
NGC~1617 & (R)SAB(s)0/a &   0                                                & NGC~7400 & SA(r\underline{s})c &   5                                     \\
NGC~1796 & SB(\underline{r}s)cd pec &   6                                    & NGC~7410 & SB$_\textrm{ax}$0$^+$ &  -1                                   \\
NGC~2082 & SA\underline{B}(rs)cd &   6                                       & NGC~7476 & (R$^{\prime}$)SA\underline{B}(rs)ab &   2                     \\
NGC~2544 & (RL)SAB(r,bl)0$^+$ &  -1                                          & NGC~7496A & SAB(s)d &   7                                                \\
NGC~2785 & I0 pec & 0                                                        & NGC~7545 & SA\underline{B}(s)dm &   8                                    \\
NGC~7713A & SAB(rs)cd &   6                                                  & PGC~052618 & SA(rs)c &   5                                               \\
PGC~001457 & SB(s)dm &   8                                                   & PGC~053303 & SA(s)c &   5                                                \\
PGC~002465 & SB(r)d &   7                                                    & PGC~053583 & S\underline{A}B(s)cd &   6                                  \\
PGC~003689 & Sd sp/ E(d)7 &   7                                              & PGC~053977 & SB(rs)dm &   8                                              \\
PGC~008762 & Im sp / E(d)8 &  10                                             & PGC~068090 & SA(s)d sp &   7                                             \\
PGC~009485 & (R)SB(rl)a &   1                                                & PGC~068107 & S0d / E(d)9 &   7                                           \\
PGC~011135 & (R$^{\prime}$)SAB(s)dm &   8                                    & PGC~069381 & SAB(s)d &   7                                               \\
PGC~014125 & SABd &   7                                                      & PGC~070104 & Scd pec &   6                                               \\
PGC~015989 & SA\underline{B}(s)dm sp / E(d)8-9 &   8                         & PGC~070154 & SB(s)d &   7                                                \\
PGC~024031 & SA(s)c: &   5                                                   & PGC~070787 & SB(s)m &   9                                                \\
PGC~025347 & dIm sp / E(d)8 &  10                                            & PGC~083842 & (RL)S\underline{A}B(l)0$^\textrm{o}$ &  -2                  \\
PGC~026260 & SA(s)d &   7                                                    & PGC~090489 & Sd sp / E(d)8 &   7                                         \\
PGC~027656 & IB(s)m &  10                                                    & PGC~091258 & Sd sp / E(d)8 &   7                                         \\
PGC~027696 & SA(s)cd &   6                                                   & PGC~1010767 & IB(s)m sp / E5-6 &  10                                     \\
PGC~028195 & S0$^-$ sp / E(d)6 &  -3                                         & PGC~1019240 & IBm sp / E7 &  10                                          \\
PGC~028253 & IB(s)m &  10                                                    & PGC~1034827 & IBm sp / E4 &  10                                          \\
PGC~030608 & Sd spw / E(d)8 &   7                                            & PGC~128826 & SA0$^-$: sp &  -3                                           \\
PGC~030906 & SA(s)c &   5                                                    & PGC~132468 & SAB(s)cd &   6                                              \\
PGC~033447 & Im? &  10                                                       & PGC~158719 & Sd spw / E(d)7 &   7                                        \\
PGC~034407 & dE4 &  11                                                       & PGC~170100 & Im sp / E7 &  10                                            \\
PGC~034625 & SB(s)d &   7                                                    & PGC~282435 & SA(s)cd &   6                                               \\
PGC~035451 & SA(s)c &   5                                                    & PGC~3097691 & dSph &  11                                                 \\
PGC~035570 & SAB(s)cd &   6                                                  & PGC~3097702 & SAB(s)dm &   8                                             \\
PGC~036904 & E(d)4 (isophote twisting)) &  -5                                & PGC~725719 & SB(s)m &   9                                                \\
PGC~037625 & SA\underline{B}(s)dm &   8                                      & PGC~891484 & (R$^{\prime}$)SA(s)d &   7                                  \\
PGC~037712 & SB(s)d &   7                                                    & PGC~896975 & SA0$^-$ sp &  -3                                            \\
PGC~039052 & SA(s)cd &   6                                                   & PGC~924727 & SAB(s)dm &   8                                              \\
PGC~039359 & SA(r)0$^+$ / E6-7 &  -1                                         & PGC~938112 & SA(s)cd sp &   6                                            \\
PGC~039748 & E3 &  -5                                                        & PGC~952909 & Im sp / E8 &  10                                            \\
PGC~040045 & SA\underline{B}(s)m &   9                                       & PGC~974148 & E6? pec &  -5                                               \\
PGC~040547 & dE & 11                                                         & PGC~975705 & SB(s)m sp &   9                                             \\
PGC~040869 & dE5,N? &  11                                                    & UGC~00685 & dIAm &  10                                                   \\
PGC~040889 & dE1,N &  11                                                     & UGC~00931 & Sd sp / E(d)8 &   7                                          \\
PGC~041314 & SAB(s)dm &   8                                                  & UGC~01050 & SAB(s)m &   9                                                \\
PGC~041387 & E6: &  -5                                                       & UGC~04783 & Sdm sp / E(d)7-8 &   8                                       \\
PGC~041463 & SAB(s)dm pec &   8                                              & UGC~04788 & (RL)SBa: &   1                                               \\
PGC~041598 & E5-6 &  -5                                                      & UGC~04832 & SB(s)0/a &   0                                               \\
PGC~042898 & Sd spw / E(d)9 &   7                                            & UGC~04879 & dIm &  10                                                    \\
PGC~042902 & E$^+$2-3 &  -4                                                  & UGC~04883 & (L)SA(l)0$^\textrm{o}$ &  -2                                 \\
PGC~043178 & E1-2 &  -5                                                      & UGC~04889 & S\underline{A}B(s)c &   5                                    \\
PGC~043283 & IABm &  10                                                      & UGC~04906 & SAb sp / E(d)8 &   3                                         \\
PGC~043526 & pec & -                                                         & UGC~04990 & SB(rs)d &   7                                                \\
PGC~043600 & SB(s)d &   7                                                    & UGC~05049 & SAb sp / E(d)8 &   3                                         \\
PGC~043690 & (R$_1^{\prime}$)SAB$_\textrm{a}$(\underline{r}l)a &   1         & UGC~05227 & SB(s)d: sp &   7                                             \\
PGC~043698 & E6 pec (2 shells?) &  -5                                        & UGC~05348 & Sd sp / E7-8 &   7                                           \\
PGC~044701 & I0 sp pec / E6 & 0                                              & UGC~05369 & Sdm sp / E(d)8 &   8                                         \\
PGC~044733 & S0c: sp &   5                                                   & UGC~05454 & (R$^{\prime}$)SA(s)d &   7                                   \\
PGC~044812 & (R$^{\prime}$)SAB(s)cd &   6                                    & UGC~05479 & SA\underline{B}(s)\underline{d}m &   7.5                     \\
PGC~044983 & (R$^{\prime}$)SB(s)c &   5                                      & UGC~05692 & dIAB(s)m &  10                                               \\
PGC~045512 & dE3,N &  11                                                     & UGC~05760 & Sb sp pec / E7 &   3                                         \\
PGC~045711 & SA(s)c &   5                                                    & UGC~05974 & SAB(rs)d &   7                                               \\
PGC~045955 & SA(rl)0/a &   0                                                 & UGC~05983 & SB(s)dm sp / E(d)7-8 &   8                                   \\
PGC~046256 & S0$^-$ sp / E(d)7-8 &  -3                                       & UGC~06061 & (R$^{\prime}$L)SB(s)d &   7                                  \\
PGC~046436 & SB$_\textrm{x}$(r)0$^+$ / E(d)6-7 &  -1                         & UGC~06102 & SB(rs)dm &   8                                               \\
PGC~046630 & Im pec / E7 &  10                                               & UGC~06448 & SB[0]d &   7                                                 \\
PGC~049341 & SB(s)dm &   8                                                   & UGC~06455 & SB(r\underline{s})cd &   6                                   \\
PGC~050551 & E4-5 pec &  -5                                                  & UGC~06611 & SB(s)dm &   8                                                \\
PGC~050651 & E(d)7? &  -5                                                    & UGC~06658 & SA(s)c: &   5                                                \\
PGC~051309 & SB(s)a &   1                                                    & UGC~06669 & IAB(s)m &  10                                                \\
UGC~06685 & S\underline{A}B(s)d sp &   7                                     & UGC~08308 & dIm &  10                                                    \\
UGC~06717 & SB(s)cd &   6                                                    & UGC~08386 & SB(s)d &   7                                                 \\
UGC~06728 & (L)SB$_\textrm{a}$(s)0$^\textrm{o}$ &  -2                        & UGC~08491 & SB(s)m / E(d)8 &   9                                         \\
UGC~06756 & I0 pec & 0                                                       & UGC~08514 & Sd: sp / E(d)8 &   7                                         \\
UGC~06764 & SAB(s)d &   7                                                    & UGC~08656 & Sd sp / E(d)8 &   7                                          \\
UGC~06881 & SAB(s)dm &   8                                                   & UGC~08731 & (R$^{\prime}$)SAB(s)d &   7                                  \\
UGC~06919 & Sbc sp pec / E(d)7 &   4                                         & UGC~08736 & SA(s)bc &   4                                                \\
UGC~06988 & Sd sp / E(d)7-8 &   7                                            & UGC~08741 & SB(\underline{r}s)cd &   6                                   \\
UGC~06992 & Im &  10                                                         & UGC~08822 & Sd sp / E(d)8 &   7                                          \\
UGC~06998 & SAB(s)cd &   6                                                   & UGC~08857 & [SA(r)0$^\textrm{o}$ / S0c(r)] sp &  -2                      \\
UGC~07038 & SB(s)dm &   8                                                    & UGC~08876 & S0$^-$ sp / E(d)6 &  -3                                      \\
UGC~07138 & SA(s)cd &   6                                                    & UGC~08970 & (R$^{\prime}$)SA(s)d &   7                                   \\
UGC~07176 & SAB(s)dm &   8                                                   & UGC~09112 & SB(s)d sp / E7-8 &   7                                       \\
UGC~07178 & SB(rs)d &   7                                                    & UGC~09251 & SAB:(s)cd sp &   6                                           \\
UGC~07179 & SAb sp / E(d)7-8 &   3                                           & UGC~09285 & Sd sp / E(d)6-7 &   7                                        \\
UGC~07238 & Sd sp /E(d)8 &   7                                               & UGC~09355 & Sd sp / E7 &   7                                             \\
UGC~07346 & dE$^+$0 &  11                                                    & UGC~09405 & dIm (dSph) &  10                                             \\
UGC~07370 & SB(s)d &   7                                                     & UGC~09432 & SAB(s)m &   9                                                \\
UGC~07392 & Im sp / E(d)8 &  10                                              & UGC~09504 & dIm &  10                                                    \\
UGC~07411 & SAB0$^-$ / E5 &  -3                                              & UGC~09637 & Sd sp / E(d)8-9 &   7                                        \\
UGC~07422 & (RL)SB(r)ab &   2                                                & UGC~09741 & (L)SB(\underline{r}s)ab &   2                                \\
UGC~07438 & I0 pec / E6-7 & 0                                                & UGC~09827 & Sdm sp / E(d)7-8 &   8                                       \\
UGC~07636 & E2 &  -5                                                         & UGC~10125 & SAB(s)d &   7                                                \\
UGC~07688 & (R$^{\prime}$)dSA(r)d &   7                                      & UGC~10369 & Sd sp / E(d)7 &   7                                          \\
UGC~07710 & IA(s)m &  10                                                     & UGC~10589 & (R$^{\prime}$)SA(s,nd)c &   5                                \\
UGC~07841 & I0 pec & 0                                                       & UGC~10625 & Sd spw /E(d)8-9 &   7                                        \\
UGC~07857 & dE4,N &  11                                                      & UGC~10892 & Sd sp /E(d)9 &   7                                           \\
UGC~07945 & SAB(s)cd &   6                                                   & UGC~10907 & S\underline{A}B(s)bc &   4                                   \\
UGC~08032 & E7 (dust lane) &  -5                                             & UGC~10974 & (R$_1$R$_2^{\prime}$)S\underline{A}B(r,nr)ab &   2           \\
UGC~08059 & Sa spw / E(d)8 &   1                                             & UGC~11964 & Sd sp / E(d)9 &   7                                          \\
UGC~08101 & SB(\underline{r}s)d &   7                                        & UGC~12300 & (L)SB$_\textrm{a}$0$^+$ &  -1                                \\
UGC~08231 & SB(s)cd &   6                                                    & UGC~12344 & Sbc: sp / E(d)8 &   4                                        \\
UGC~08245 & dIm (dSph) &  10                                                 & UGC~12729 & I0 pec & 0                                                   \\

\end{longtable}

\end{appendix}
\end{document}